\documentclass{elsart}

\usepackage{epsfig}
\usepackage{multirow}

\usepackage[nofiglist,notablist,nomarkers]{endfloat}

\newcommand{\ic}{\'{\i}}

\begin{document}
%\addtolength{\baselineskip}{0.5cm}

\begin{frontmatter}
\title{ A possible use for polarizers in Imaging Atmospheric Cherenkov
Telescopes } 
\thanks[corresp]{Corresponding author.    E-mail address: contrera@gae.ucm.es}
\author[leeds]{I. de la Calle},
\author[madrid]{J.L. Contreras\thanksref{corresp}},
\author[munich]{J. Cortina},
\author[madrid]{V. Fonseca}
\address[leeds]{ University of Leeds Department of Physics and Astronomy, LS2 9JT Leeds, UK }
\address[madrid]{Universidad Complutense, 
Facultad de Ciencias F\ic sicas, Ciudad Universitaria,
Avda Complutense s/n E-28040 Madrid, Spain }
\address[munich]{Max Planck Institut f\"ur Physik, 
F\"ohringer Ring 6, D-80805, M\"unchen, Germany }
\begin{abstract}
 Cherenkov radiation produced in Extensive Air
Showers shows a net polarization. This article discusses its properties
and physical origin, and proposes an arrangement of polarizers potentially 
useful for Imaging Atmospheric Cherenkov Telescopes.
\end{abstract}
\end{frontmatter}

\newpage

\section{Introduction}
\label{sect1}

The purpose of this work is twofold. For one, it studies a
widely neglected 
aspect of Very High Energy (VHE) Extensive Atmospheric 
Air Showers (EAS), namely the 
polarization of the Cherenkov light. 
For another, it proposes an experimental
setup that could be used to take advantage of the net polarization in EAS
in the field of Imaging Atmospheric Cherenkov Telescopes (IACTs).
 
We shall proceed as follows. To begin with, the
basic physical ideas that explain why a net polarization of 
the Cherenkov light is expected in an EAS will be
outlined and some features of this polarization will be
discussed. Section \ref{sect3} will propose an experimental setup 
which exploits
the net intrinsic polarization of showers initiated by $\gamma$-rays and in a
lesser degree of those initiated by charged cosmic rays. 
A detailed study of the setup has been
carried out using a Monte Carlo (MC) simulation that will be described in
section \ref{sect4}. The result of this study is presented in
section \ref{sect5}. A short discussion relevant to some related situations
not considered in the paper has been left for section \ref{sect6}.

\section{Physical basis}
\label{sect2}

\begin{figure}[p]
\begin{center}
\epsfig{file=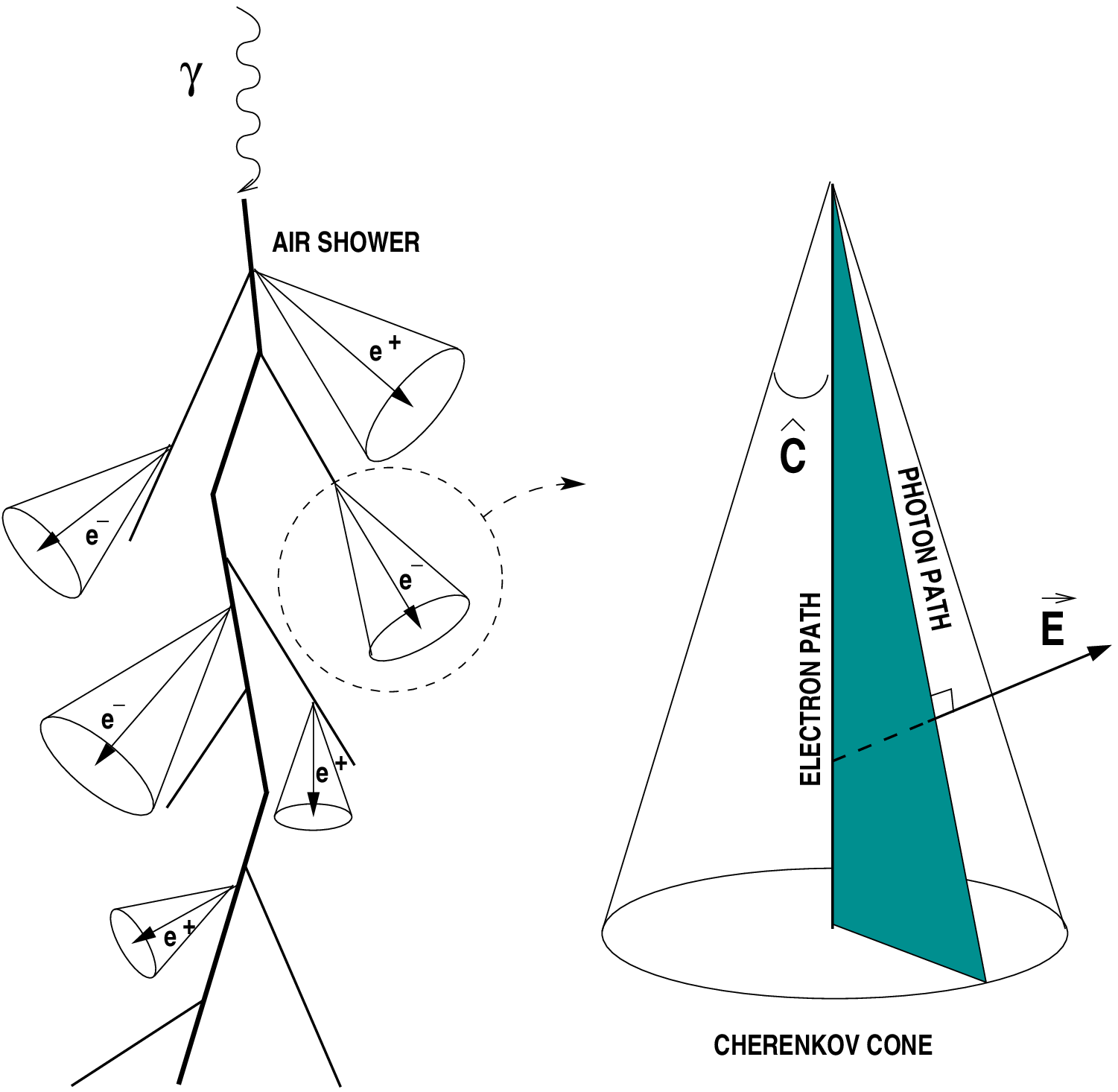,height=8cm,width=.8\linewidth}\\
\epsfig{file=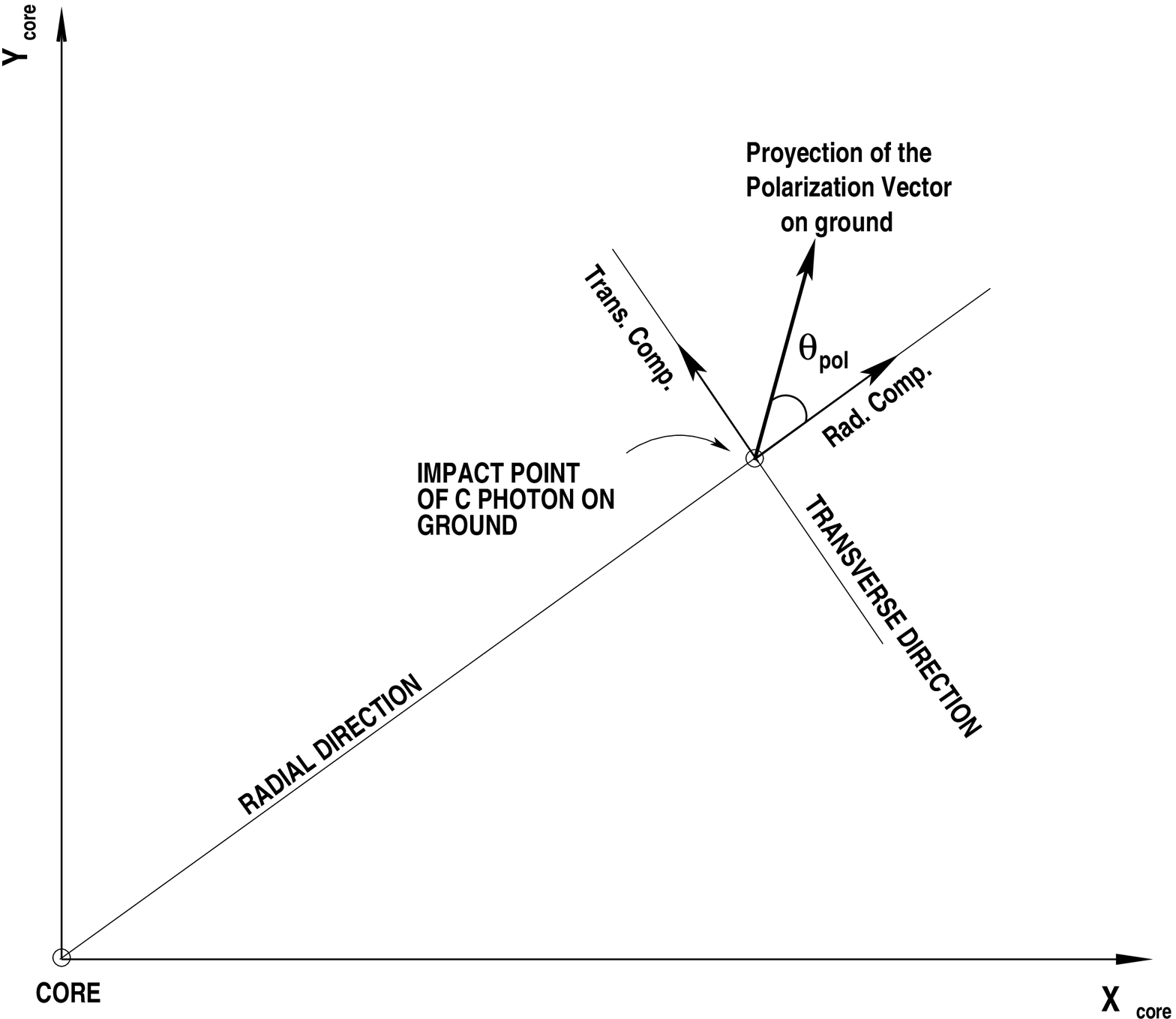,width=.48\linewidth}%
\epsfig{file=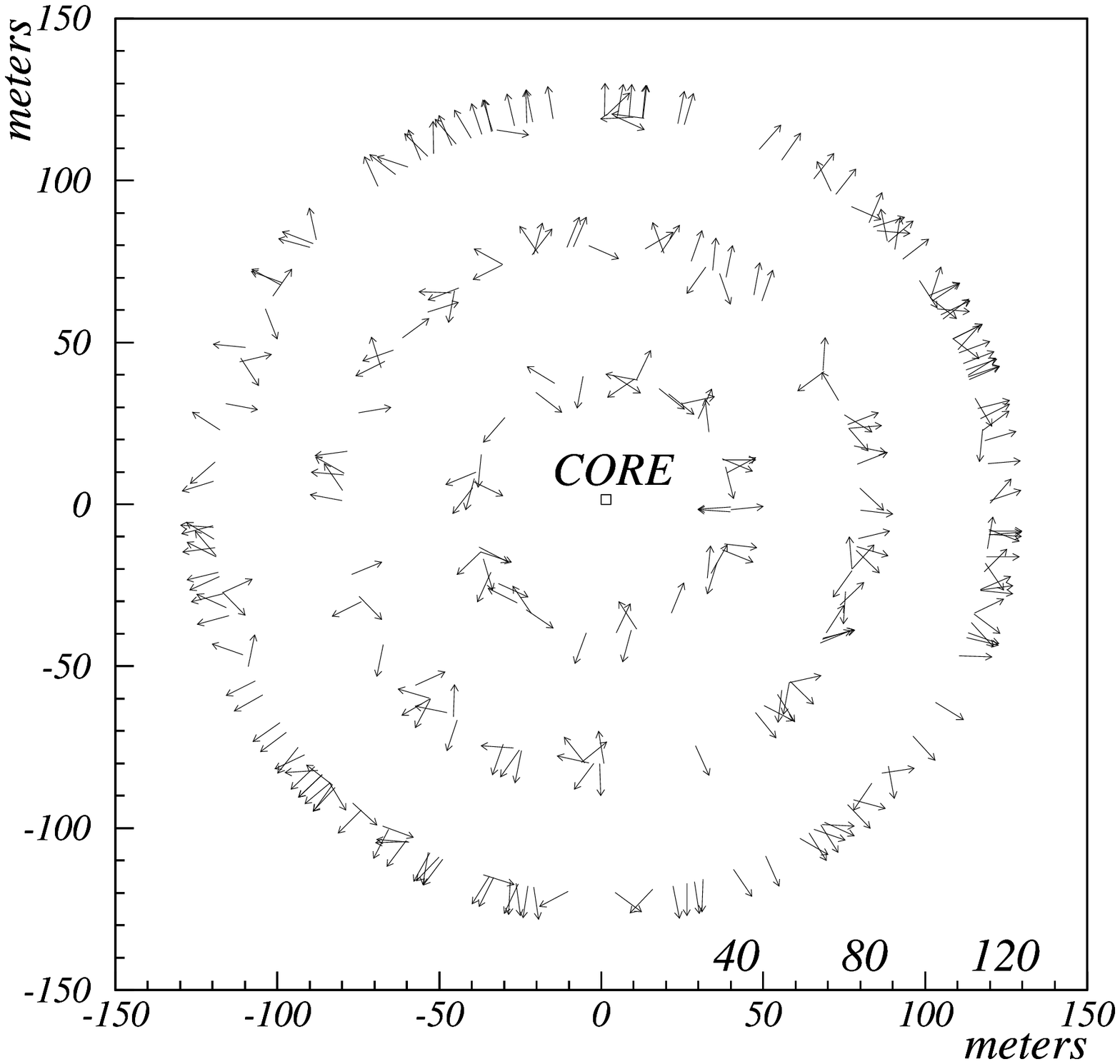,width=.48\linewidth}
\caption{\it Top left: a schematic representation of an air shower illustrating
single particle cherenkov cones. Top right: 
the polarization vector is contained
in the plane defined by the electron path and the photon path and is
perpendicular to the photon path. Bottom left, the coordinate system and
polarization parameters as referred in the text are illustrated. Bottom right,
polarization vector distribution on ground for a 1 TeV $\gamma$-induced 
MC event at three core distances. Each circle represents all
the Cherenkov photons falling at that particular core distance.}
\label{fig1}
\end{center}
\end{figure}

The Cherenkov radiation emitted by a relativistic charged particle
moving in a dense medium is distributed along a
cone whose apex moves with the particle and whose axis is
the particle path.  The opening angle of the Cherenkov cone in the 
air is of the order of one
degree. It is well known that Cherenkov radiation exhibits an
intrinsic polarization~\cite[J. V. Jelley 1958]{Jelley}. The electric 
field vector of the emitted light is contained in the plane defined by the
particle and photon paths (see Fig. \ref{fig1} upper right).

The existence of a net linear polarization in the light produced by an 
EAS can be readily understood within a simple limiting case. Suppose a
normally incident shower with respect to the ground,  whose particles
travel along the shower axis.
The light pattern left on the ground by the shower particles would be a 
superposition of circles centered in the shower core,
defined as the point where the shower axis touches ground. The 
polarization of the photons hitting the ground 
would always lie in the line passing through their point of impact on ground
and the shower core.

To describe the depicted situation it is convenient to define a
system of cylindrical coordinates, centered at the core, with the 
$Z$ axis directed upwards along the shower axis.
In this frame the polarization
of the Cherenkov photons lie in the $XY$ plane, in lines pointing to 
the core. The expression {\it radial direction} will often be used to mean
the direction of these lines. While the polarization of the 
Cherenkov photons in a real shower does not always follow 
the {\it radial direction}, it falls
to a good approximation on the $XY$ plane.  
Therefore the angle which they form with the {\it radial direction}
is a good measure of their deviation from the limiting case
described before.
Hereafter it will be named $\theta_{pol}$.

A real shower is not expected to behave in a manner substantially
different to the limit case described before.
A net polarization in the radial direction can be expected 
when the distance of the incident point to the core is much larger 
than the distance of the emitting
particle to the shower axis (i.e. when the shower is observed with less
transverse angular spread) and the direction of the emitting 
particle is similar to the shower axis. 
Both conditions are satisfied in a wide range of
situations, since the density of particles in an EAS is maximal at small
angles and distances with respect to the shower core. 
Factors like the Coulomb scattering and the hadronic interactions
are determinant to disperse the shower particles in 
angle and core distance and thus are expected to
influence the degree of net polarization along the 
radial direction. 
  
As an illustration Fig. \ref{fig1}, at the right bottom,
shows the distribution of polarization
vectors at three different core distances for a
MC-generated shower initiated by a $\gamma$-ray.
 
%----------------------------------
%    Figura  1 
%----------------------------------

Besides the above-discussed general considerations, the polarization of the
Cherenkov light carries additional information about the
EAS. This will be discussed in more detail in the following sections. The
accompanying plots have been generated using a MC simulation whose
setup is described in section \ref{sect4}. Throughout all the 
sections it will be assumed that incident $\gamma-ray$ showers
are normal to the ground,
i.e. the zenith angle is equal zero. This assumption is not as 
restrictive as it
may seem, since the reflector of an IACT is always practically
perpendicular to the shower axis, a situation in many ways equivalent 
to normal incidence.

In our simulation we have not considered the correlation between the degree of
polarization and the wavelength of the incoming Cherenkov photons. Its study
calls for a detailed model of the atmosphere and the polarizing filter.
Nevertheless the different light attenuation in the atmosphere
at different wavelengths must lead to a noticeable correlation between
these two quantities.
For instance, the observed ultraviolet light is mostly produced in
the final stages of shower development since it is strongly attenuated
in the atmosphere.
It must therefore
present a lower degree of polarization than light at other wavelengths. This
could prove interesting for experiments operating at these wavelengths 
such as CLUE ~\cite[B. Bartoli 1998]{Clue}.

\subsection{Dependence on the distance to the core and particle type} 
\label{subsect21}

%----------------------------------
%    Figura  2
%----------------------------------

Showers initiated by $\gamma$s are closer to the aforementioned 
ideal shower than hadron-initiated showers.

The reason lies in the relatively large transverse momentum of
the hadronic interactions compared with the purely electromagnetic 
interactions in gamma-ray showers.  This results in a general spread
in the shower particles.
 Therefore the net polarization 
along the radial direction is expected to be
larger in $\gamma$ showers than in hadronic ones, as already noted by Hillas
~\cite[A. M. Hillas 1996]{Hillas}. Clear differences are seen in the radial
distributions of both types of primary particles.

At first sight, the general arguments discussed above seem to imply an
increase of the degree of radial polarization with core
distance {\it r}. The reason being that the region around the 
shower axis with significant particle density is observed 
under an always smaller angle as we move away from the shower core
in the ground. However a detailed simulation shows that the degree of radial
polarization is not a monotonically increasing function of 
{\it r} (as illustrated in Fig. \ref{gamma}). Whereas the polarization
behaves as expected at {\it r} out to the so-called {\it hump} 
of the lateral distribution ($\sim$ 120 m), other
factors, such as multiple scattering, reduce the 
degree of polarization at larger {\it r}.
The result is a maximum in the lateral distribution of the degree 
of polarization.

\begin{figure}[p]
\begin{center}
\epsfig{file=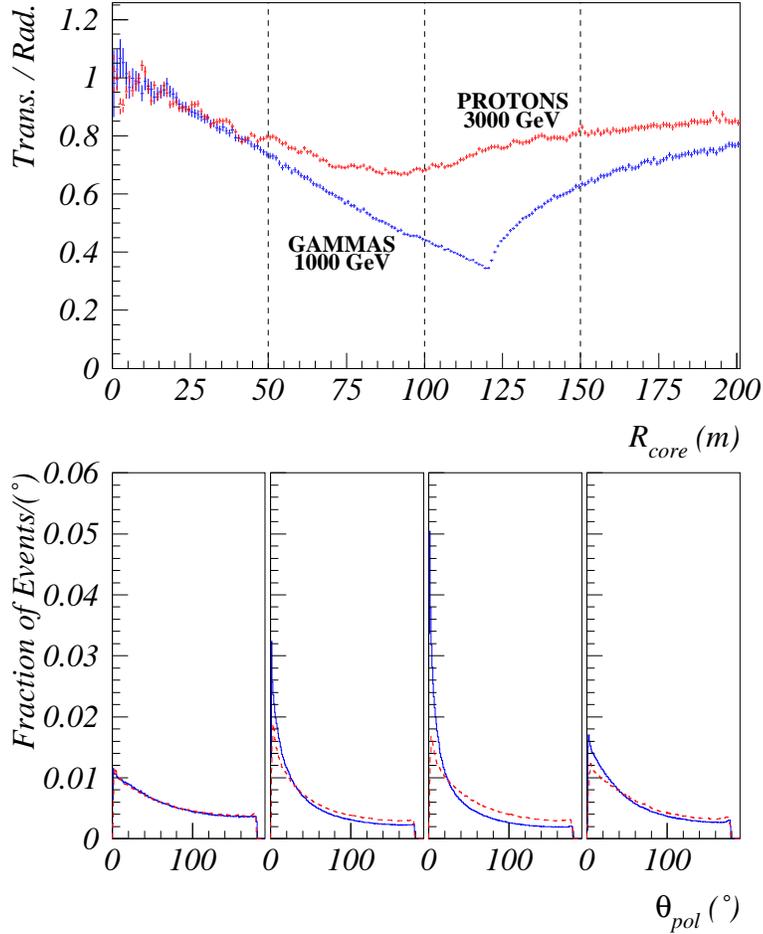,width=.8\linewidth}\\
\caption{\it On the upper plot the transverse component over the radial
component of the polarization vectors is represented as a function of the core
distance for a superposition of 10 $\gamma$ MC events at 1 TeV and a
superposition of 10 hadron MC events at 3 TeV within 3 degrees of
zenith angle . On the lower plots the distribution of the angle
$\theta_{pol}$ over four different core distance intervals, the ones indicated
by the vertical dashed lines on the top plot, is shown. The dashed line on the
bottom plots stands for protons.}
\label{fig2}
\end{center}
\end{figure}

\begin{figure}[p]
\begin{center}
\epsfig{file=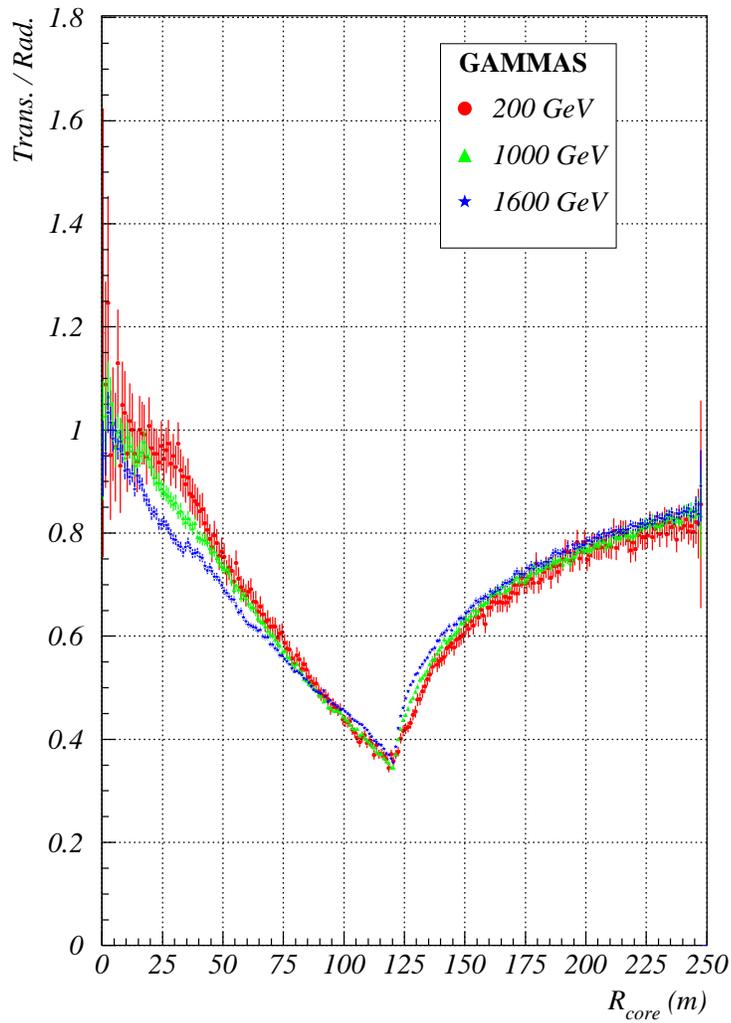,width=.8\linewidth}
\caption{\it MC simulation of the polarization distribution at the ground
for primary $\gamma$s of different energies. The distribution shown is the
same as in figure \ref{fig2} where each curve is a superposition of 10
showers.}
\label{gamma}
\end{center}
\end{figure}

The behavior of the degree of polarization with respect to {\it r} 
depends on
the nature of the particle which gives origin to the EAS. Showers
originated by Very High Energy $\gamma$-rays display a maximum 
around the {\it hump} of the Cherenkov lateral distribution. 
EAS of hadronic origin exhibit a not so abrupt turning point
at shorter {\it r} (as can be seen in Fig. \ref{fig2}). 
The influence of the primary particle species was
already noted by Hillas in ~\cite[A. M. Hillas 1996]{Hillas}
and advanced as a $\gamma$/hadron discriminator.

%----------------------------------
%    Figure  4
%----------------------------------

In addition there is a small dependence with the energy of the primary
particle (see Fig. \ref{gamma}). The 
position of the maximum is however energy independent.

Although no attempt has been made to understand in detail
the difference with particle species, physical arguments
point to the influence of multiple scattering as determinant of the
position of the maximum, and transverse momentum
to the difference between hadrons and $\gamma$s. In the case of $\gamma$
EAS, this belief is enforced by the fact that the maximum is reached near the
{\it hump}, the point which would mark the edge of the Cherenkov pool if no
multiple scattering was present. Other correlated factors, as the detailed
shape of the transverse profile may also influence this result.

\vskip 1cm
\subsection{Dependence on emission point} 
\label{subsect22}

%----------------------------------
%    Figure  5
%----------------------------------
The above-depicted limiting case (all particles moving along the shower axis)
is closer to the trajectories of the particles in the first 
stages of a real shower development than in its later
stages.         
The reason is that multiple scattering, and transverse 
momentum in the case of  hadronic interactions, have not diffused
the shower particles yet. As a result the
Cherenkov light emitted by particles in the first stages of shower development
shows a higher degree of polarization.

\begin{figure}[p]
\begin{center}
\epsfig{file=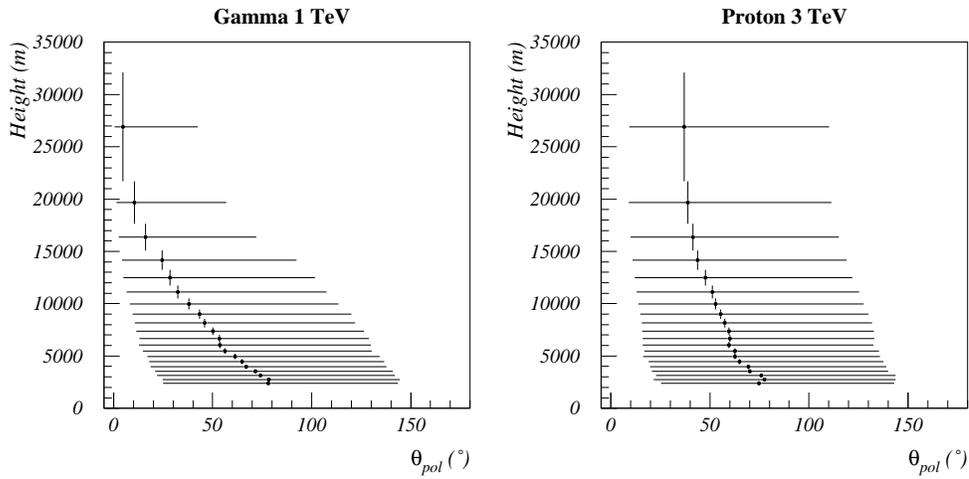,width=\linewidth}
\caption{\it Cherenkov light production height as a function of
polarization angle for 
10 MC gamma ray showers at 1 TeV and
10 MC hadron showers of 3 TeV at 3 degrees zenith angle.
The first interaction point was fixed at 35 km above sea level.
The vertical error bars represent the height interval where the
mean value of $\theta_{pol}$ has been obtained.
Since the distribution of $\theta_{pol}$ is asymmetric, 
a left and right width have been calculated 
representing the interval within 34 $\%$ of the events (equivalent to 0.5
$\sigma$ in a gaussian distribution) are contained at both 
sides of the median.}
\label{altura}
\end{center}
\end{figure}

Fig. \ref{altura} shows the median of the
$\theta_{pol}$ distribution, in a plane perpendicular to the shower axis,
as a function of the atmospheric 
depth at which the Cherenkov photons have been produced. This median
has been computed in bins of one radiation
length. Horizontal bars represent the 
dispersion of the distribution of $\theta_pol$,  
computed as the interval containing
68\% of the distribution, half of it at each side of the median.
To produce the plot, ten
gamma and ten proton showers have been generated. The first
interaction point has been fixed at 35 km. The figure shows how the median 
and the spread of $\theta_{pol}$ increase as the shower develops. 
It indicates a decrease of the degree of polarization along the
radial direction with atmospheric depth, both for $\gamma$ and proton
initiated EAS.

\subsection{Correlation of polarization and time of arrival} 
\label{subsect23}

%----------------------------------
%    Figure  6
%----------------------------------
There is also a strong correlation between the time of arrival of the
Cherenkov photons at ground and the value of their radial polarization. 
For particles traveling along the shower axis it
can be easily shown that light
at a core distance {\it r} can only arise from a limited range of
heights. Therefore the time of arrival of the photons
is also expected to be in a limited range.  
As most of the
particles travel close to the shower axis, this range defines the time of
arrival of the Cherenkov front. 
Light arriving at different times comes
from regions of the shower far from the shower axis. 
It follows that the set
of particles producing the light with the highest degree of polarization also
produces the photons which define the shower front. 
Selecting photons whose
polarization vector is aligned to the radial direction should 
then produce a narrower light pulse.
Through this paragraph the expresion {\it shower front} must
be understood as the arrival time of the peak of the light
distribution, not as the arrival time of the first photon. This 
definition generally corresponds to the experimentally measured 
magnitude.

\begin{figure}[p]
\begin{center}
\epsfig{file=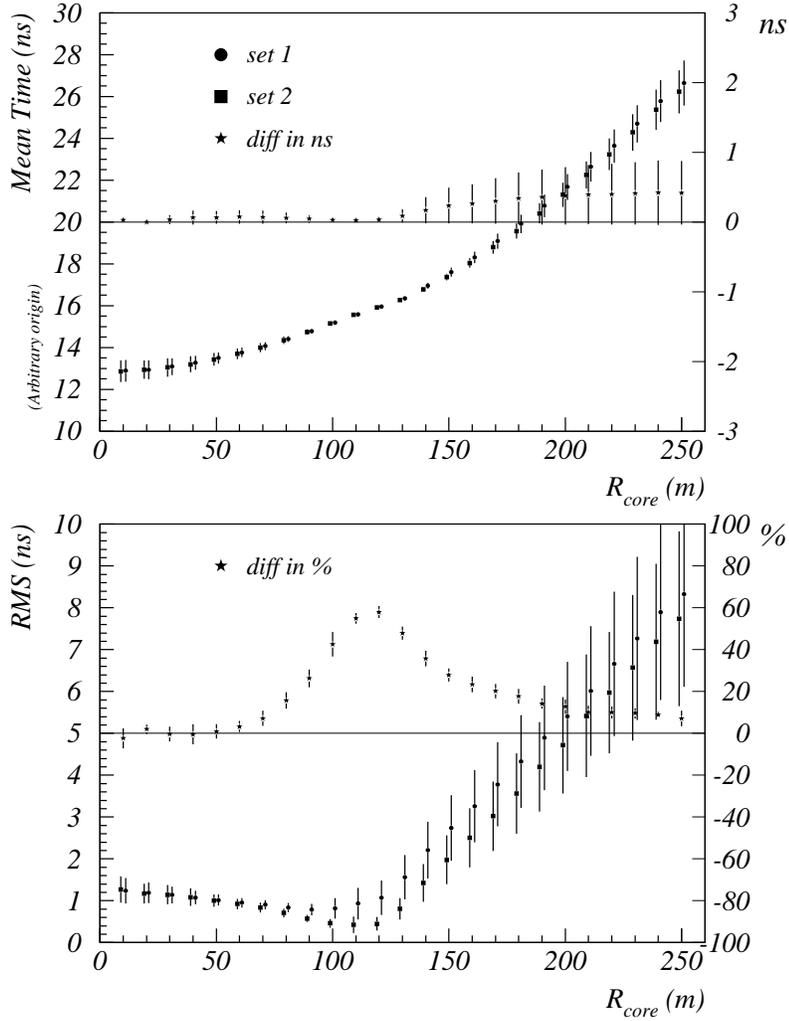,width=.8\linewidth}
\caption{\it The upper plot shows the arrival mean time at 
different core distances for the set of MC events of
the previous figure.
%1 TeV $\gamma$ MC event at zero degrees zenith angle with a first interaction
%point fixed at 35 km above sea level used before.
The time origin of the y-axis is arbitrary. 
The lower plot displays the rms of the time distribution at different core
distances. Set 1 corresponds to the polarization assumption 
and set 2 to the non-polarization assumption. 
Stars represent the difference between both
sets, expressed in nanoseconds in the upper plot and as a
percentage in the lower one. The plots are only meant
to illustrate the general features of the arrival time distribution.}
\label{tiempo}
\end{center}
\end{figure}

This is illustrated in Fig. \ref{tiempo}, which shows
the mean arrival time of the light front as a function of {\it r},
the distance to the core in the transverse plane, for 
a set of  1 TeV MC $\gamma$-ray showers. 
The distributions for all the light and for the 50\% of the light whose 
polarization vector is closer to the radial direction 
($\theta_{pol} < 32^\circ$) are compared. 
The width of the time distribution is again defined as the interval
containing the 68\% of the events, 34\% at each side of the median. 
By doing so we identify the RMS of the time distribution with the decay
time of the Cherenkov pulse. 
This decay time is more sensitive to the primary species as
noted in Ref. \cite[V. R. Chitnis and P. N. Bhat 1999]{Chitnis}.  
From this comparison we can draw the following conclusions:

\begin{itemize} 

\item The value of the mean arrival time of the
Cherenkov light front at different core distances does not depend on the
polarization distribution on ground.  In other words the use of polarizers
does not appreciably distort the light front.  

\item The width of the Cherenkov
light pulse depends on the polarization distribution on ground. The difference
in percentage between both cases is maximum close to the hump.  

\end{itemize}
    
We have not performed an exhaustive and quantitative study of 
the magnitude of these effects at other primary energies, but general 
arguments suggest that both conclusions will hold valid 
for a large range of energies.

 The advantage of working with smaller pulse widths stems 
from the reduction it induces in the total amount of NSB.
In the usual situation the significance of a
Cherenkov signal over the
background will approximately be of:$ \frac{Signal}{\sqrt{NSB}}$.
 In turn , the NSB factor is directly proportional to the
time gate during which the ADC system integrates light.
Therefore a reduction factor $F$ in the width of the ADC
gates,  which does not affect the signal, translates 
aproximately in an increase of $\sqrt{F}$ in the significance
for Cherenkov signals.
 The above defined significance
is related to the threshold for both gamma and hadronic 
showers with respect to a fixed level of NSB and should not be confused
with the quality factor of the gamma/hadron separation cut.

\section{Application to IACTs }
\label{sect3}

As shown in previous sections, the Cherenkov light 
at ground is significantly polarized along the
radial direction, defined by
the shower core and the observation point.
This fact may be useful in attempting to  
determine the position of the shower core
(as in Ref. \cite[A. K. Tickoo 1999]{Tickoo}),
but may be seen as a drawback to exploit the
potential for gamma/hadron discrimination
of the polarization, since in most cases
the core position is ignored {\it a priori}.
However, for the particular case of IACTs the
characteristics of the imaging apparatus 
allow to make use of this potential
by using a simple experimental setup. 

\par

Let us firstly recall that light does not 
loose its state of polarization 
after normal reflection in a mirror. 
Cherenkov light of showers triggering an
IACT reaches the mirrors almost perpendicularly. 
This owes to the fact that the opening angle of
the telescope is of the order of a few degrees
and that Cherenkov photons travel almost parallel
to the direction of the shower axis.
Therefore polarization information is 
conserved upon reflection on the mirror.

 It must also be remembered that atmospheric showers
give rise to elliptical images on the camera,
where the major axis of the ellipse points 
to the direction of the shower core. 
For showers arising from the direction
the telescope is pointing to,
the major axis of the ellipse will lay along the line
which joins the center of the camera and the 
shower core. This happens to be the direction along
which the Cherenkov photons which form the image are polarized.
The argument boils down to saying that
the {\it radial} polarization on the ground 
translates into a {\it radial}
polarization on the camera for showers coming in
the direction of the telescope axis.

%----------------------------------
%    Figure  7
%----------------------------------

\begin{figure}[p]
\begin{center}
\epsfig{file=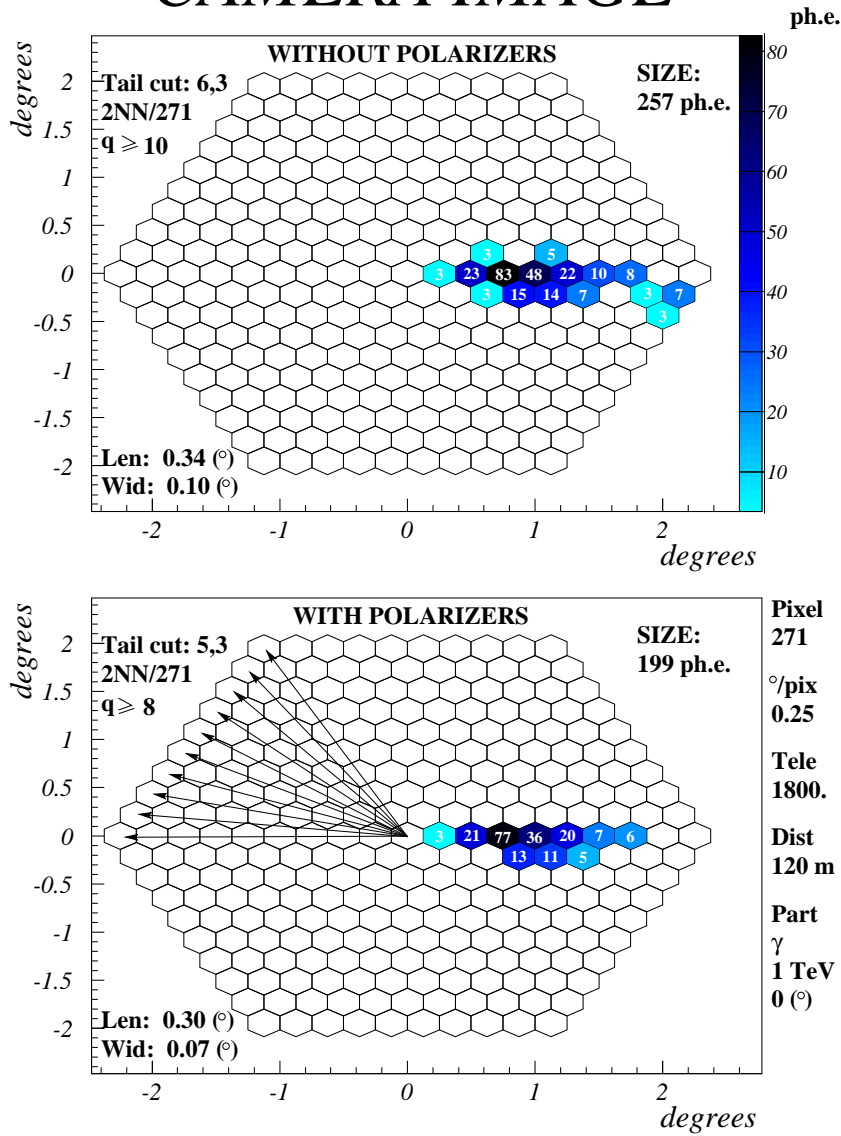,width=.8\linewidth}
\caption{\it 
The upper plot shows the image produced in the standard
camera by a MC 1 TeV $\gamma$-ray shower. 
The lower plot shows the image of the same shower 
in the polarizer camera described on 
the text. The arrows in the lower plot
illustrate how the polarizers are arranged on the camera.}
\label{setup}
\end{center}
\end{figure}

Fig. \ref{setup} shows the schematics of the setup which
we shall use to take advantage of the radial polarization of
the light in the image. 
A polarizer is
laid on top of each of the camera photomultipliers (PMTs) with 
its transmitting axis pointing towards the center of the camera.
The arrows in the lower figure indicate
the direction of the polarizer axes.
In the following we shall refer to the camera equipped
with polarizers as {\it polarizer camera} as opposed to
the {\it standard camera} without polarizers.
The signal in the pixels of both cameras
by a simulated $\gamma$-ray event 
is also displayed in the figure.

It is well known that the projection of the light density 
onto the ellipse major axis is correlated to
the longitudinal development of the shower. 
Pixels closest to the camera
center correspond to early stages in the shower development and 
those farthest away to its latest phases. 
The degree of radial polarization should thus increase
slightly towards the center of the telescope where 
light comes mainly from
higher up in the atmosphere.

%----------------------------------
%    Figure  7
%----------------------------------

\begin{figure}
\begin{minipage}[p]{.50\linewidth}
\centering\epsfig{file=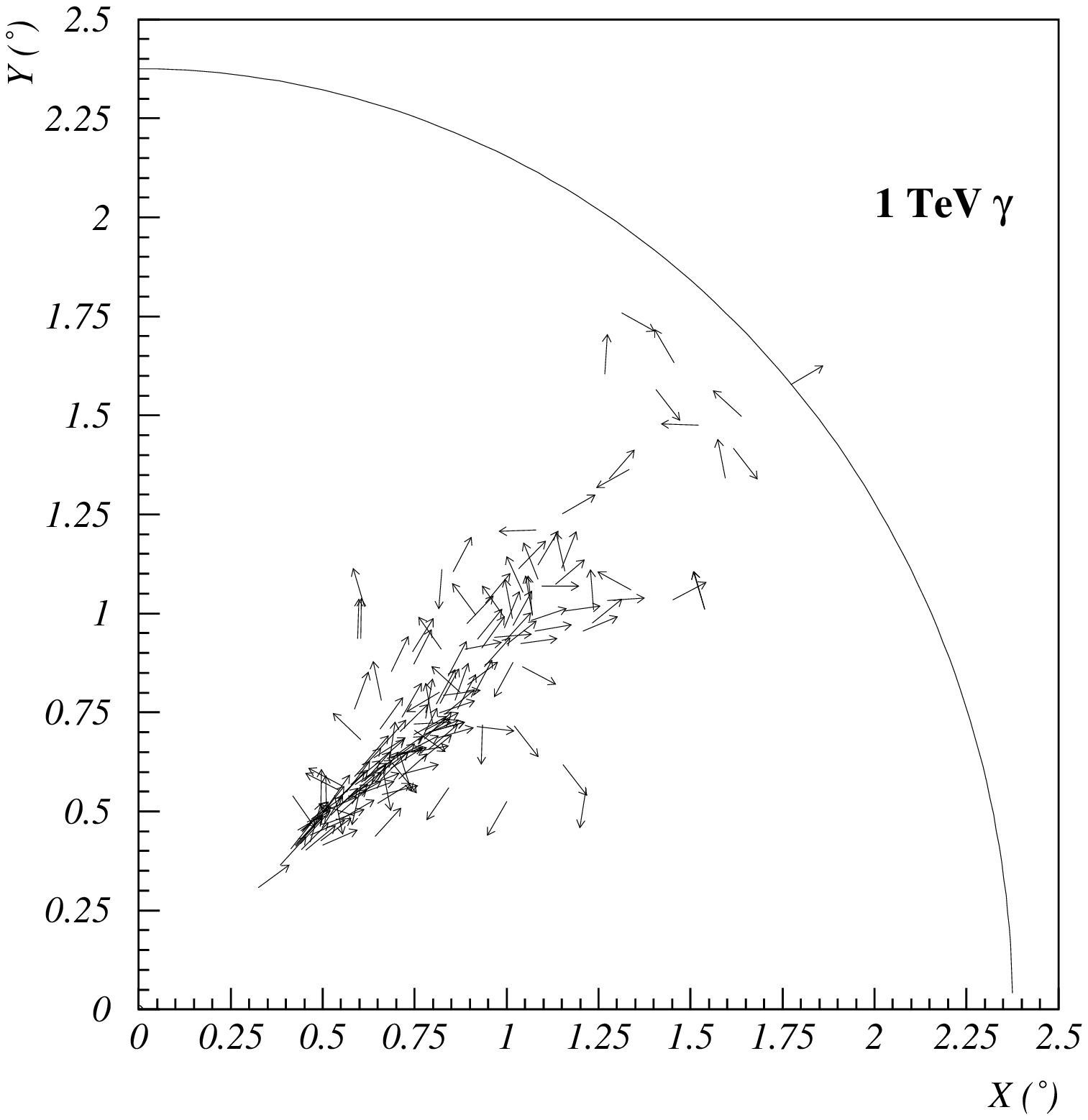,height=5.6cm,width=.95\linewidth}
\end{minipage}\hfil
\begin{minipage}[p]{.50\linewidth}
\centering\epsfig{file=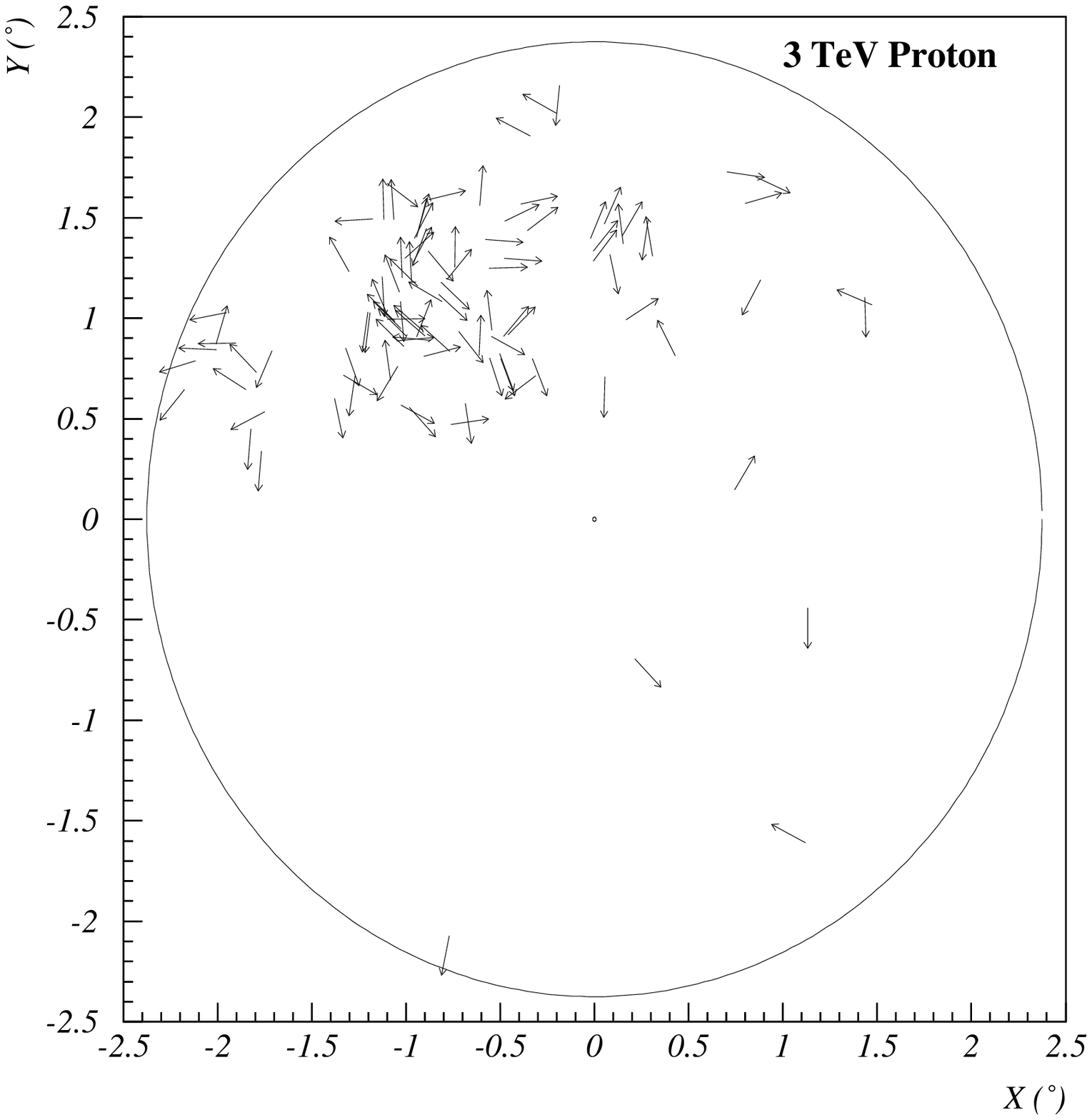,height=5.6cm,width=.95\linewidth}
\end{minipage}\hfil
\caption{\it MC simulation of the polarization vector distribution on
the IACT focal plane. {\bf On the left:} 1 TeV gamma event of impact parameter
close to the hump. Only one fourth of the camera is shown, 
the rest is empty. 
{\bf On the right:} 3 TeV proton event at $3^\circ$ zenith angle and impact
parameter close to the hump.}
\label{flechas}
\end{figure}

Let us see how this setup may enhance a $\gamma$-ray signal in an IACT.
Since night sky light is unpolarized,
an immediate effect of this setup is to suppress 50\% 
of the background light.
In addition
the images of off-axis showers are distributed 
randomly over the camera, which is to say that major axes
do not point to the center of the camera.
Since most of the hadron showers are off-axis, we expect
the setup to reduce the hadronic background.
Besides, the light in the tail and outer zones of the image 
will be substantially reduced and the image will be compacted. 
This is because the radial component of the polarization 
dominates close to the camera center.
Fig. \ref{flechas} illustrates this point for
a simulated gamma and a simulated proton shower.

\section{Monte Carlo Simulation}
\label{sect4}

To study the proposed setup we have used a MC 
simulation based on the program CORSIKA 5.6 
\cite[D. Heck et al. 1998]{Corsika}, slightly modified it to
include the polarization of the Cherenkov light. 
We have selected the packages VENUS \cite[K. Werner 1993]{Venus}
and GHEISHA \cite[H. Fesefeldt 1985]{Gheisha} to correspondingly
simulate high and low energy hadronic interactions
and EGS4 ~\cite[W. R. Nelson et al. 1985]{Egs4}
to simulate the electromagnetic interactions.

100 $\gamma$-initiated showers were generated at
discrete primary energies in the range 200 GeV - 1.6 TeV
in steps of 100 GeV and 100 proton-initiated 
showers at primary energies in the range 600 GeV - 4.8 TeV
in steps of 300 GeV.
The difference in the energy range of the simulated
showers obeys to the fact that showers induced by
$\gamma$-rays produce roughly three times more Cherenkov 
light that showers induced by protons.
The telescope was always simulated as pointing to the zenith.
Gamma showers were produced only in the zenith direction while 
protons were distributed isotropically within 3$^\circ$
of the zenith.
Observation level and magnetic field correspond to
the HEGRA experiment site at El Roque de los Muchachos 
($28^\circ$N, $17^\circ$E, 2200 meters a.s.l.) in the Canary Islands.

\subsection{Detector Simulation} 
\label{subsect42}

We shall simulate an IACT with the features
of the telescope in the HEGRA system 
\cite[A. Daum et al. 1997]{System}
but apply a number of simplifications.
The reflector is represented
by a single 8.5 m$^2$  parabolic mirror
of 5 m focal length. The camera is located
at the focal point and consists of 271 
0.25$^\circ$ diameter PMTs.
We assume that a global $10\%$ photon conversion 
efficiency for the whole detector accounts
for atmospheric attenuation, mirror reflectivity
and QE of the PMTs.

In order to increase the statistics of our sample 
we image each individual EAS with $\sim 10^4$
telescopes homogeneously arranged inside a circle 
of 250 meters radius centered on the shower core.
This procedure leads to a sample on the order of $10^6$ 
gamma events and $10^6$ proton events for each primary 
energy. Whilst the sampling fluctuations are faithfully
represented, the sample suffers from a high shower to shower 
correlation.

We suppose that the polarizers are perfect, that is, they
absorb no light polarized in the transmitting axis. 
The effect of a camera with and and without polarizers 
has been studied for each event.
We also assume that the polarization vector suffers 
no change in the state of polarization upon reflection 
on the mirror. This should be a good approximation. 
We refer the reader to Ref. 
\cite[J. Sanchez Almeida and V. Martinez Pillet 1992]{Polar} 
for a complete discussion on the effects that
a telescope has on polarized light.

The simulation also takes into account the effect 
of the night sky background (NSB) by assuming an intensity of 
$(1.7 \pm 0.4) \cdot 10^{12}$ ph m$^{-2}$ sr$^{-1}$ s$^{-1}$.
This number draws on experimental measurements
at El Roque de los Muchachos \cite[R. Mirzoyan and E. Lorenz]{NSB}
using a narrow angle detector in the 300-600 nm
spectral range.

\subsection{Trigger Condition} 
\label{subsect43}

A critical factor in the IACT technique is the trigger 
threshold condition.
The trigger threshold sets the limit to 
the accidental trigger rate mainly due to NSB. 
The polarizing filters reduce the NSB reaching
the camera PMTs in a factor two,
hence reducing the rate of accidental triggers.
Conversely the trigger threshold must be different
to accomplish the same rate of accidentals for
a camera with and without polarizers.
A trigger condition of two next neighbor
pixels (NN) with at least 10 photoelectrons (phe)
(as currently in use in the HEGRA system
\cite[A. Konopelko  1999]{Konopelko})
was adopted for the simulation of the camera with no polarizers.
The demand to obtain the same rate of accidental
triggers leads to a trigger condition of two NN with
at least 8 phe in the case of camera
equipped with polarizers.

We shall concern ourselves with two different background conditions. 
To start with we shall deal with 
NSB corresponding to a dark night, i.e., with no moon present.
Then we will consider a higher NSB condition 
as during  twilight or in the presence of moonlight. 
In the latter case the NSB can grow up to 
a factor of 50 ~\cite[D. Kranich et al. 1999]{Kranich}
with respect to a moon-less night.
In our simulation we have adopted an intermediate
ten-fold factor increase.

In making moonlight observations
different approaches have been followed, such as 
to insert filters in front of the PMTs \cite{moonwhipple}
or to reduce the gain of the PMTs. 
We shall compare simplified versions of these techniques with the
above-described arrangement of polarizers.
Of course inserting filters and polarizers 
is a one step process and can not be strictly
compared to changing PMT gain in an arbitrary value.
Polarizers and filters may however be
considered as a first step in reducing the NSB.

A different trigger condition was defined for each NSB condition
and experimental setup. 
The trigger conditions are obtained by  
requiring the same probability of random triggers.
They are tabulated in table \ref{Tab1}.

%-------------------------------------------
%
%   Table 1
%
%------------------------------------------- 

In obtaining the effective area and energy threshold of the
telescope, we have assumed a Crab-like $\gamma$-ray 
source of differential spectral index 2.6 and flux
normalization factor
$dJ_\gamma$ ($>$1 TeV) = 2.8 $\cdot$ 10$^{-11}$ s$^{-1}$ cm$^{-2}$ TeV$^{-1}$.
Background protons have been simulated assuming
a differential index 2.7 and flux normalization factor
$dJ_{CR}$ ($>$1 TeV) = 
1.6 $\cdot$ 10$^{-5}$ s$^{-1}$ cm$^{-2}$ sr$^{-1}$ TeV$^{-1}$.
Only background events between $0^\circ$ and
$3^\circ$ zenith angle have been considered.

\subsection{Image Analysis} 
\label{subsect44}

Cherenkov images are generally described in terms of the so-called
Hillas image parameters ~\cite[A.M. Hillas 1985]{Jolla}
(a second moment analysis of the
intensity distribution of light in the camera).
A two level so-called tail-cut is applied
to extract the shower image from the light intensity on the
pixels (as described e.g. in Ref. \cite[M. Punch et al. 1992]{Punch}). 
We have selected the values of the so-called picture and boundary 
tail-cuts to be above a certain
threshold imposed by the fluctuations of the NSB. 
For the low NSB situation with no polarizers the
image tail-cut matches the cut applied in the HEGRA experiment
(6 phe for the picture tail-cut and 3 phe for the boundary 
tail-cut \cite[A. Konopelko 1999]{Konopelko}).
When simulating the polarizer camera both cuts were
tuned to obtain approximately the same probability
of random occurrence and turned out to be 5 phe 
for the picture tail-cut and 3 phe for the boundary tail-cut.

Background-dominated pixels can be excluded by ways of this technique.
Only then are image parameters computed.
We have applied a set of super-cuts in agreement with those 
used by the HEGRA collaboration
(see ~\cite[D. Petry et al. 1996]{mrk421}) and kept it constant 
over all the camera arrangements under discussion.
This means that some room is still left open to optimize the cuts
in the polarizer camera.

\section{Results}
\label{sect5}

%----------------------------------
%    Figure  8
%----------------------------------

The goal of this section is to estimate the performance of 
the proposed polarizer arrangement. 
We shall consider a detection and an analysis level.

\par

In the detection level we shall compare the effective area of the
telescope ($A_{eff}$) for $\gamma$- and hadron-initiated EAS. 
$A_{eff}$ is computed as:

\begin{center}
\begin{equation}
A_{eff,\gamma} (E) = 2 \pi \int_{0}^{\infty} P_{\gamma}(E,r) \cdot r dr
\end{equation}
\end{center}

where $P_{\gamma}(E,r)$ is the trigger efficiency or probability of detecting
a gamma ray shower of primary energy $E$ at core distance $r$. For
hadron-initiated EAS the integral the angle of incidence is also considered.

Each MC shower allows us to estimate the
effective area as the total area covered by telescopes, multiplied by the
fraction of the total number of simulated telescope
positions for which the simulated telescope response gives
raise to a trigger.

\begin{center}
\begin{equation}
A_{eff,i} (E) = A_{tot} \times \frac{N_{tel,triggered}}{N_{tot}}
\end{equation}
\end{center}
  
We calculate $A_{eff}$(E) as the average of $A_{eff,i}$(E) 
for all the simulated showers.

The differential $\gamma$ detection rate is obtained
by weighing $A_{eff,\gamma}$ with the spectrum of the incident Very High Energy
$\gamma$s. The threshold energy, $E_{th}$, is defined 
as the value of the energy for which the rate of
detection of $\gamma$ showers is maximal.
Although $E_{th}$ contains essentially
the same information as $A_{eff}$, it has a more appealing
physical meaning and it is easier to compare with
experimental results and other simulations.

At the analysis level, a standard analysis based on the Hillas parameters
\cite[A. M. Hillas 1985]{Jolla} will be performed on both samples. On one
hand the distributions of the most significant image parameters give some
insight at the differences induced by the presence of the polarizers. On the
other hand the quality factor obtained from the analysis will also be
compared.

The two different scenarios of respectively low and high NSB described in
section \ref{subsect43} will be separately analyzed and compared.

% For each of
%them the essential results of the simulation performed will be presented in a
%set of figures. Each of them shows the comparison of the behavior of one of
%the parameters studied between the {\it normal} case, when no polarizers are
%placed in front of the camera, and our proposed setup.
 
\subsection{Low noise scenario}
\label{subsect52}

\begin{figure}[p] 
\begin{center}
\epsfig{file=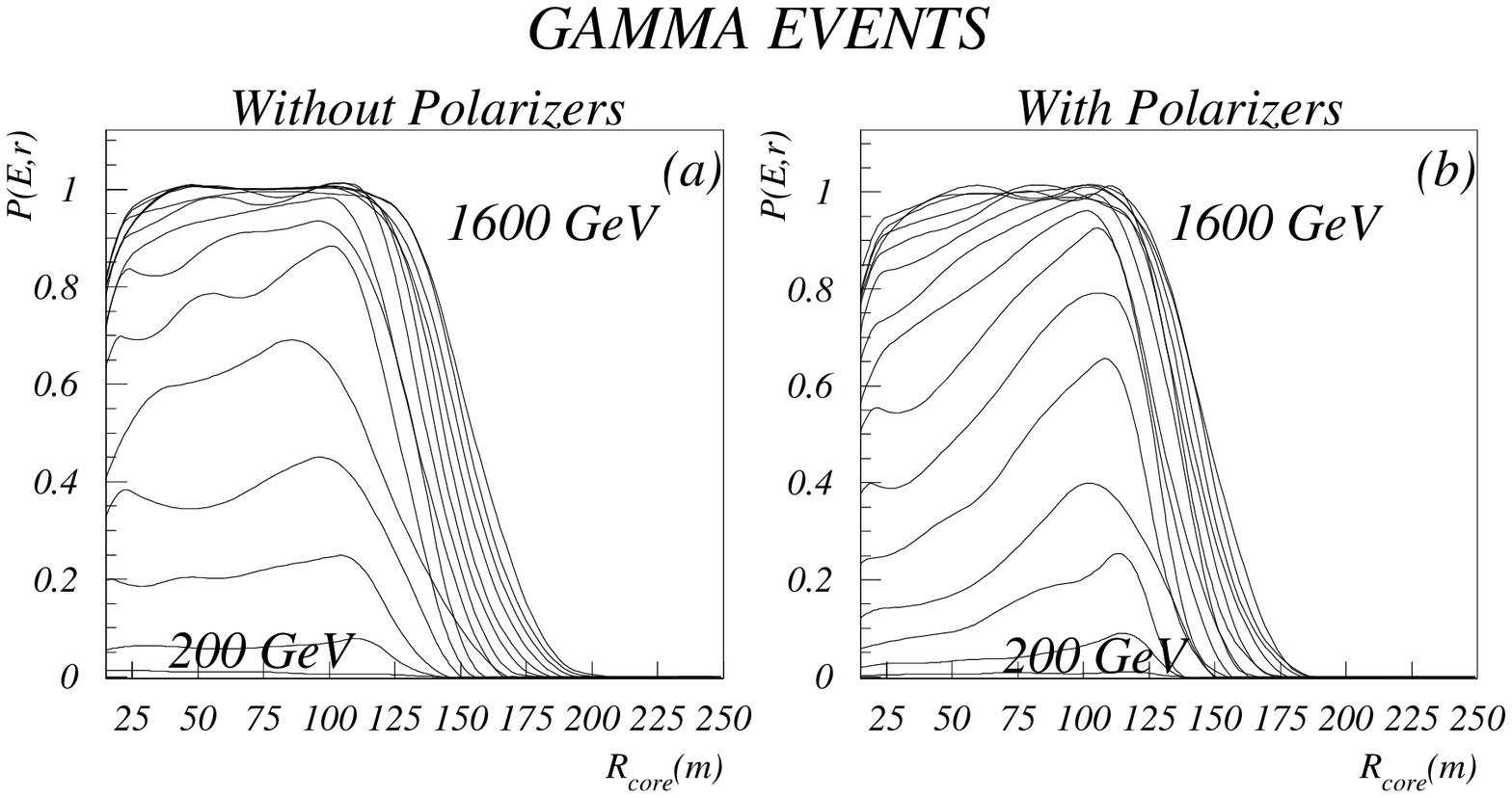, width=.9\linewidth}\\
\epsfig{file=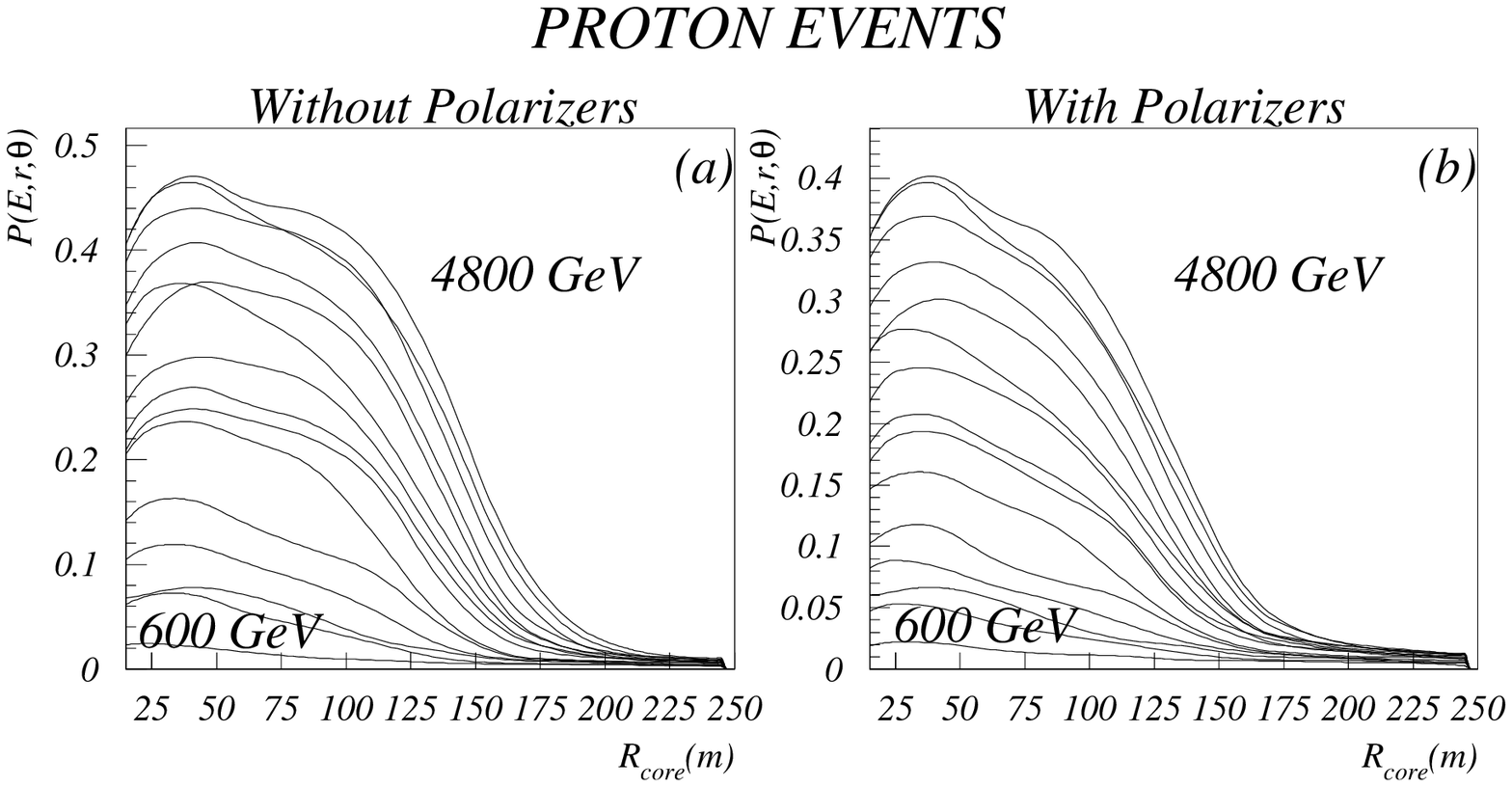, width=.9\linewidth}
\caption{\it The upper plot represents the detection probability 
as a function of primary $\gamma$ energy for a standard ({\bf a})
and a polarizer camera ({\bf b}). The lower figure shows the same
for proton events.}\label{probabilities}
\end{center} 
\end{figure}

Fig. \ref{probabilities} shows the trigger probability for $\gamma$ and
proton initiated events. For $\gamma$ events differences due to
the introduction of the polarizers are more observable at low energies and
close to the core, as expected from the total amount of light collected and
the distribution of net polarization. For proton events 
the effect of the polarizers is more noticeable in the
total number of triggered events than in the 
shape of the trigger distribution.

%----------------------------------
%    Figure  9
%----------------------------------

\begin{figure}[p] 
\begin{center}
\epsfig{file=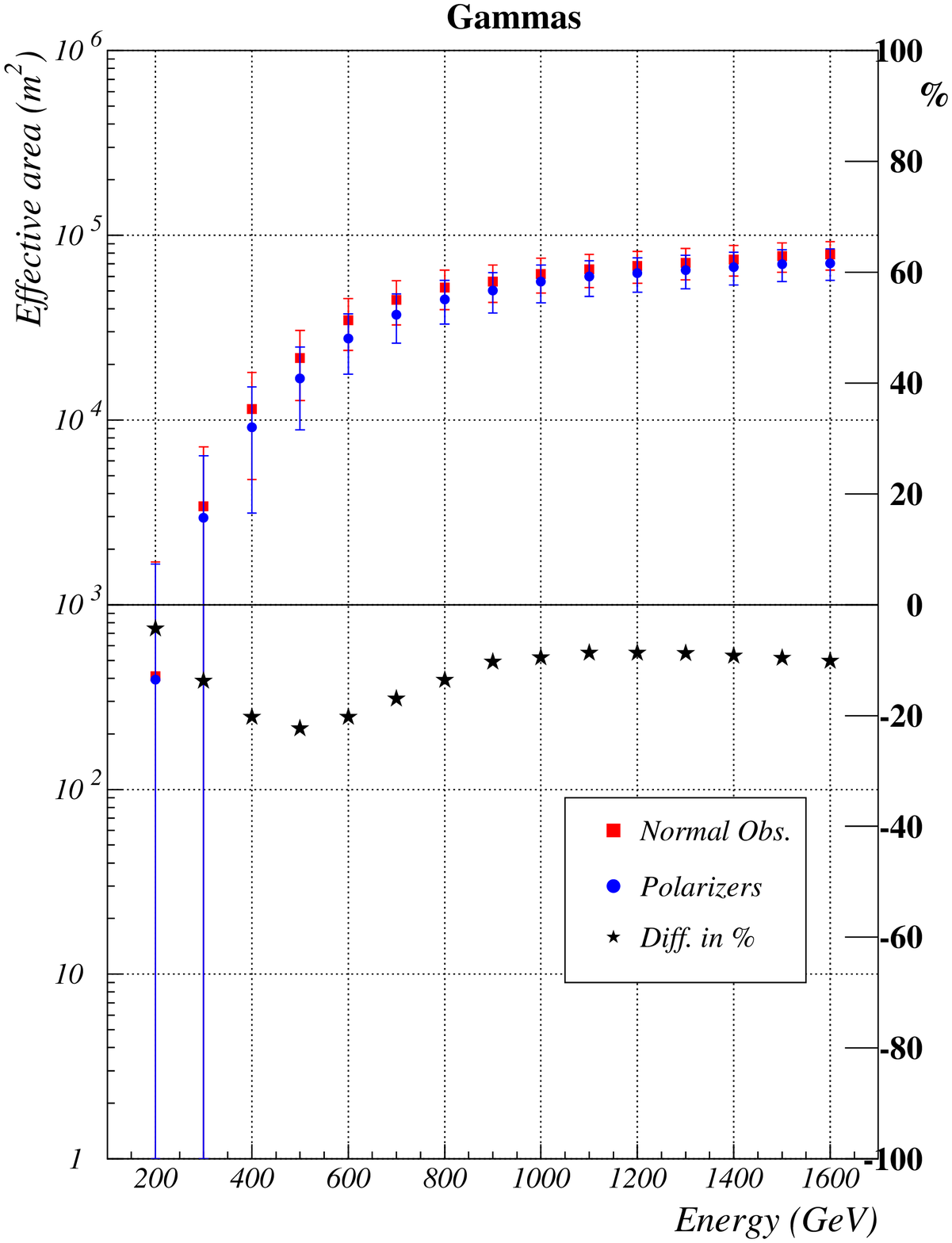,height=8cm,width=0.48\linewidth}%
\epsfig{file=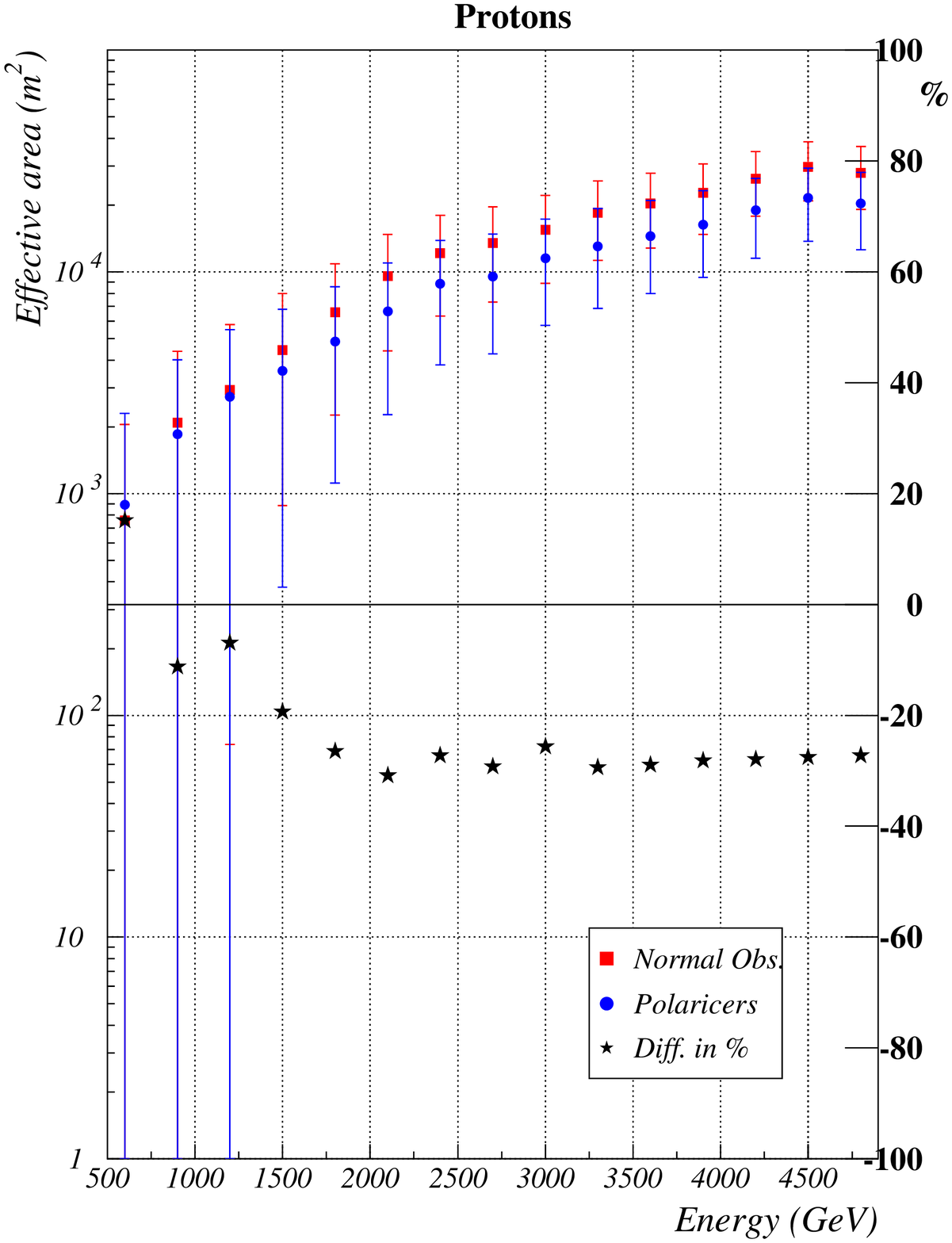,height=8cm,width=0.48\linewidth}\\
\caption{\it Shown here are the effective areas for $\gamma$ and 
proton showers for a standard and a polarizer camera. 
The right scale measures the percentage difference between the areas 
in both cameras.}\label{aeflow}
\end{center} 
\end{figure}

Fig. \ref{aeflow}  shows the effective area for $\gamma$ and proton initiated
events. Two facts must be noted:

\begin{itemize} 

\item $A_{eff}$(E) is always smaller when polarizers are used, 
both for $\gamma$ and protons.  

\item The loss in effective area is larger for protons than for $\gamma$s.

\end{itemize}

 The conclusion can be drawn that although the background is
reduced more strongly than the signal , this improvement
does not compensate for the lost in statistics due to the
reduction in triggers. The net reduction in the signal to
noise ratio between the two cameras is of approximately
10\%.

%The conclusion can be drawn that 
%although the signal to noise ratio improves (as indicated in 
%the second conclusion), this improvement does not suffice 
%to compensate for the reduction in triggers.
%Nevertheless the difference between the two setups is only of 
%the order of 10\%.

%----------------------------------
%    Figure  10
%----------------------------------

\begin{figure}[p] 
\begin{center}
\epsfig{file=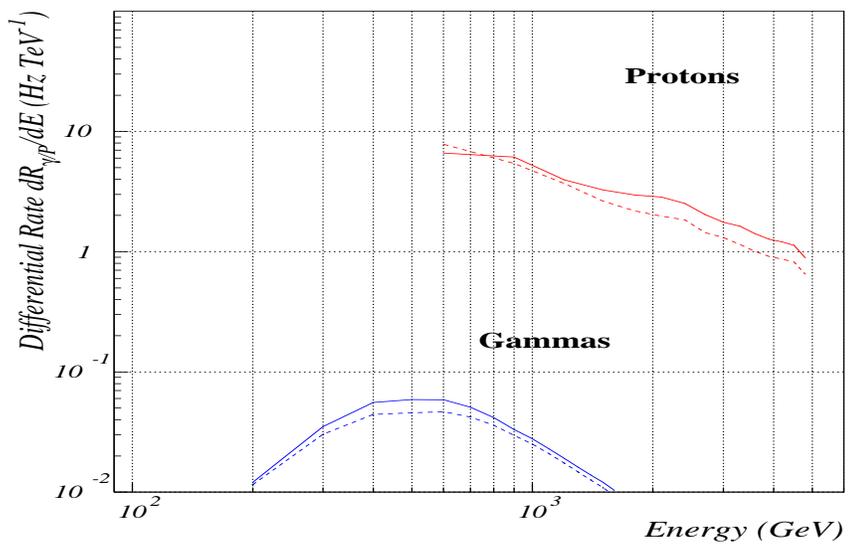, height=8cm, width=.9\linewidth}\\
\caption{\it Differential rates as a function of energy
for the standard camera (solid lines) and the polarizer camera
(dashed lines).}\label{difrate}
\end{center} 
\end{figure}

Fig. \ref{difrate} shows the differential rate as a function of
primary energy for $\gamma$ and proton initiated events. The energy 
threshold at around 500 GeV compares reasonably
well to the results in Ref. \cite[A. Konopelko 1999]{Konopelko} based in a
detailed simulation of the HEGRA telescope system and matched to 
experimental data. Additionally, it is seen that the
introduction of polarizers implies no significant difference
in energy threshold but a small reduction of the effective area.

\begin{figure}[p]
\begin{center}
\epsfig{file=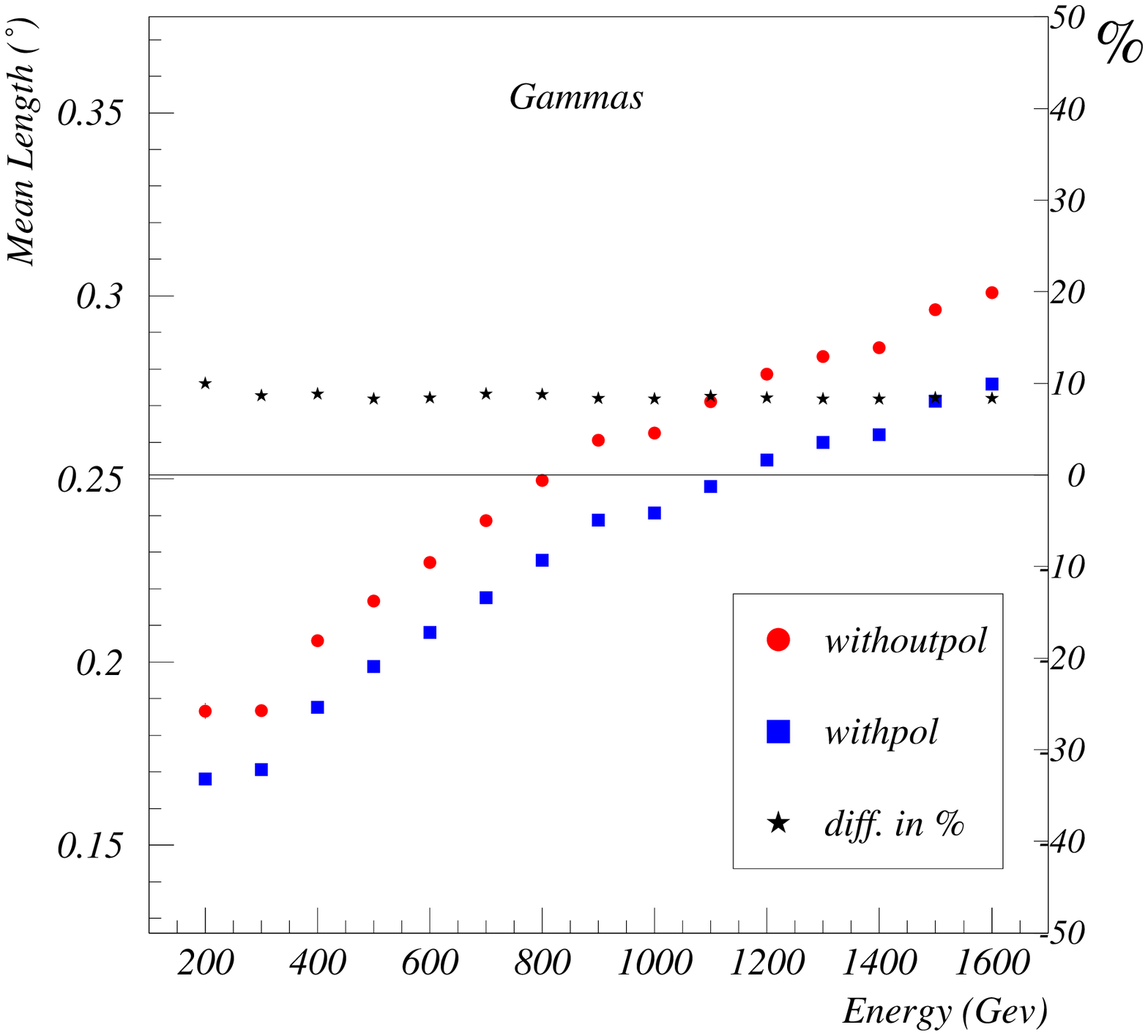,height=6.5cm,width=0.49\linewidth}%
\epsfig{file=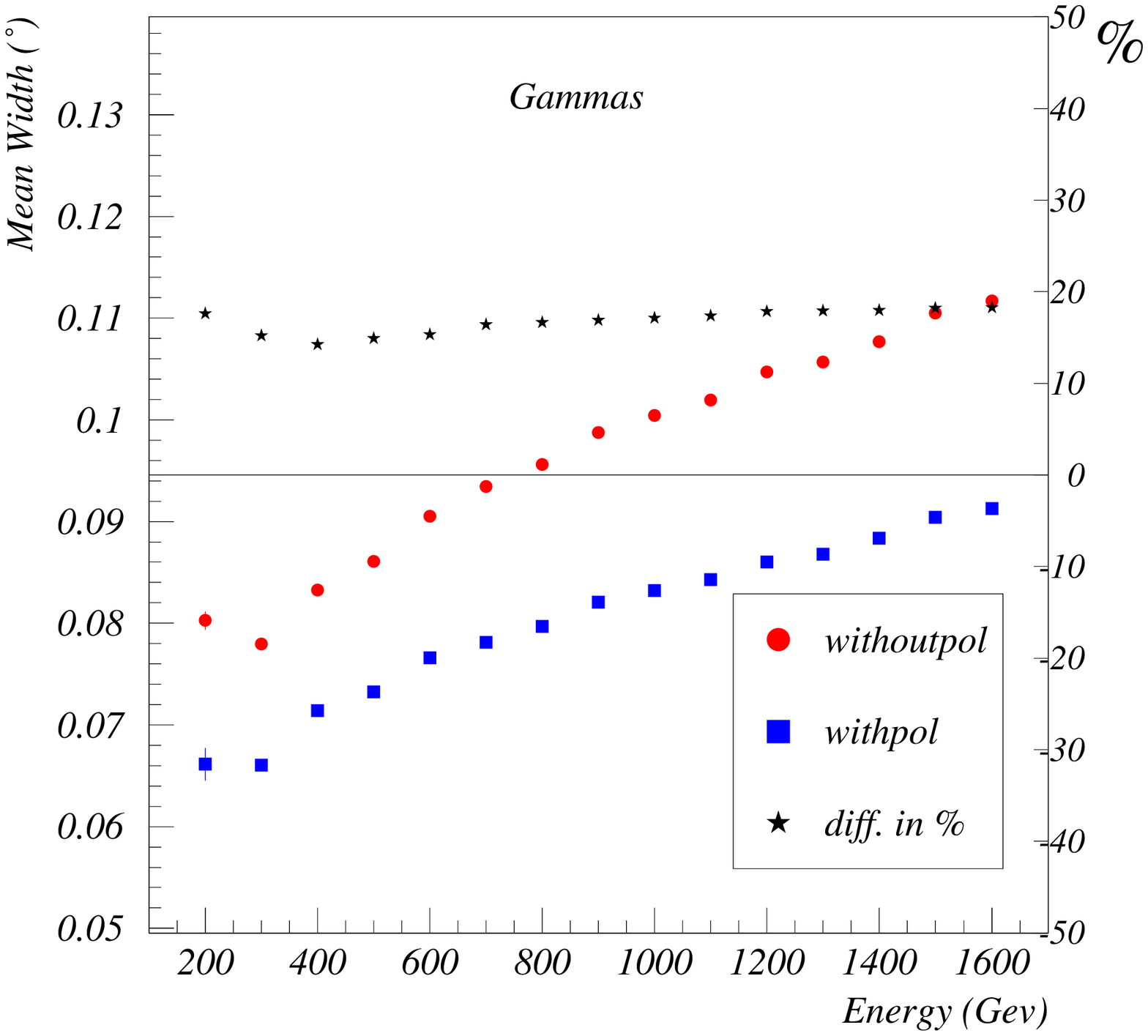,height=6.5cm,width=0.49\linewidth}\\
\epsfig{file=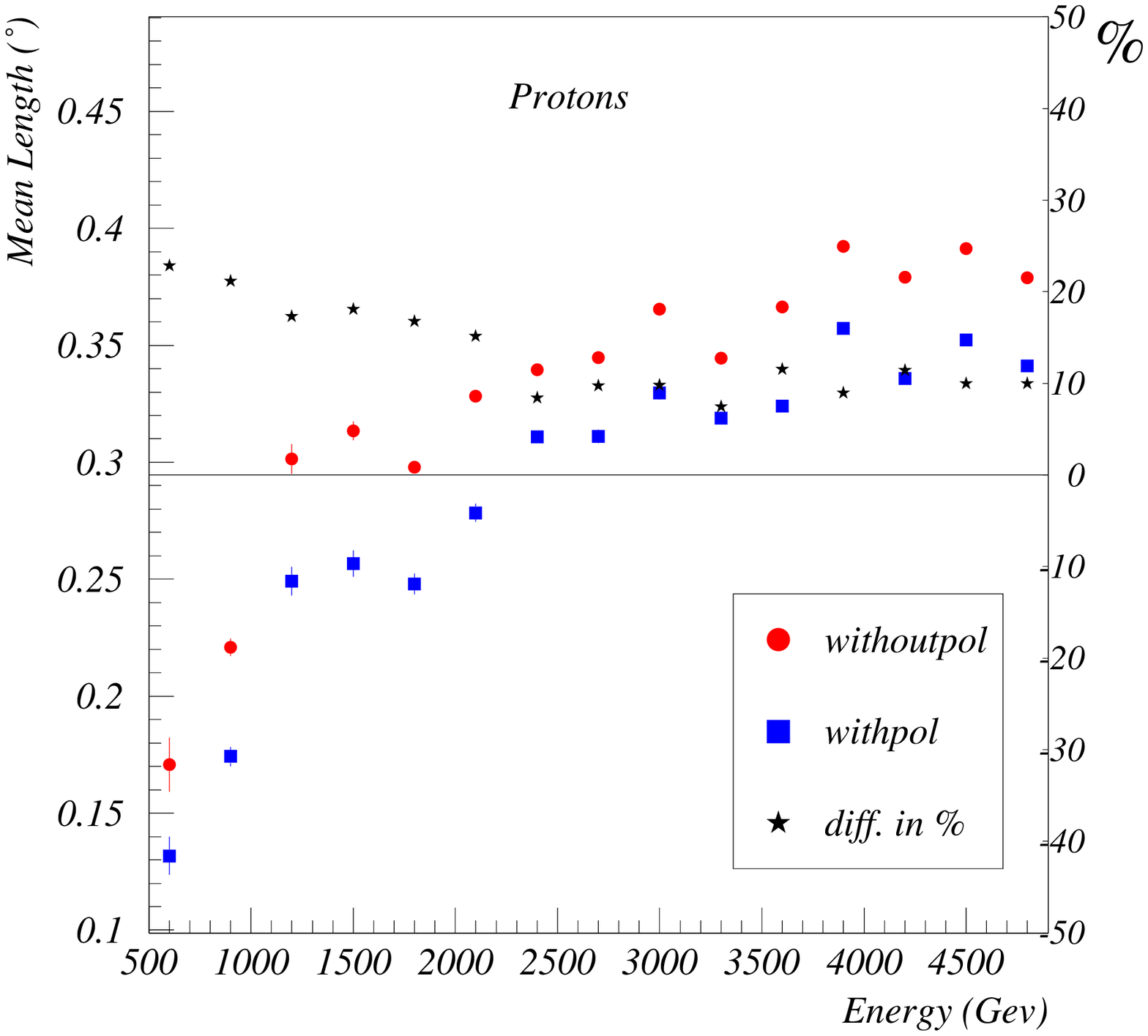,height=6.5cm,width=0.49\linewidth}%
\epsfig{file=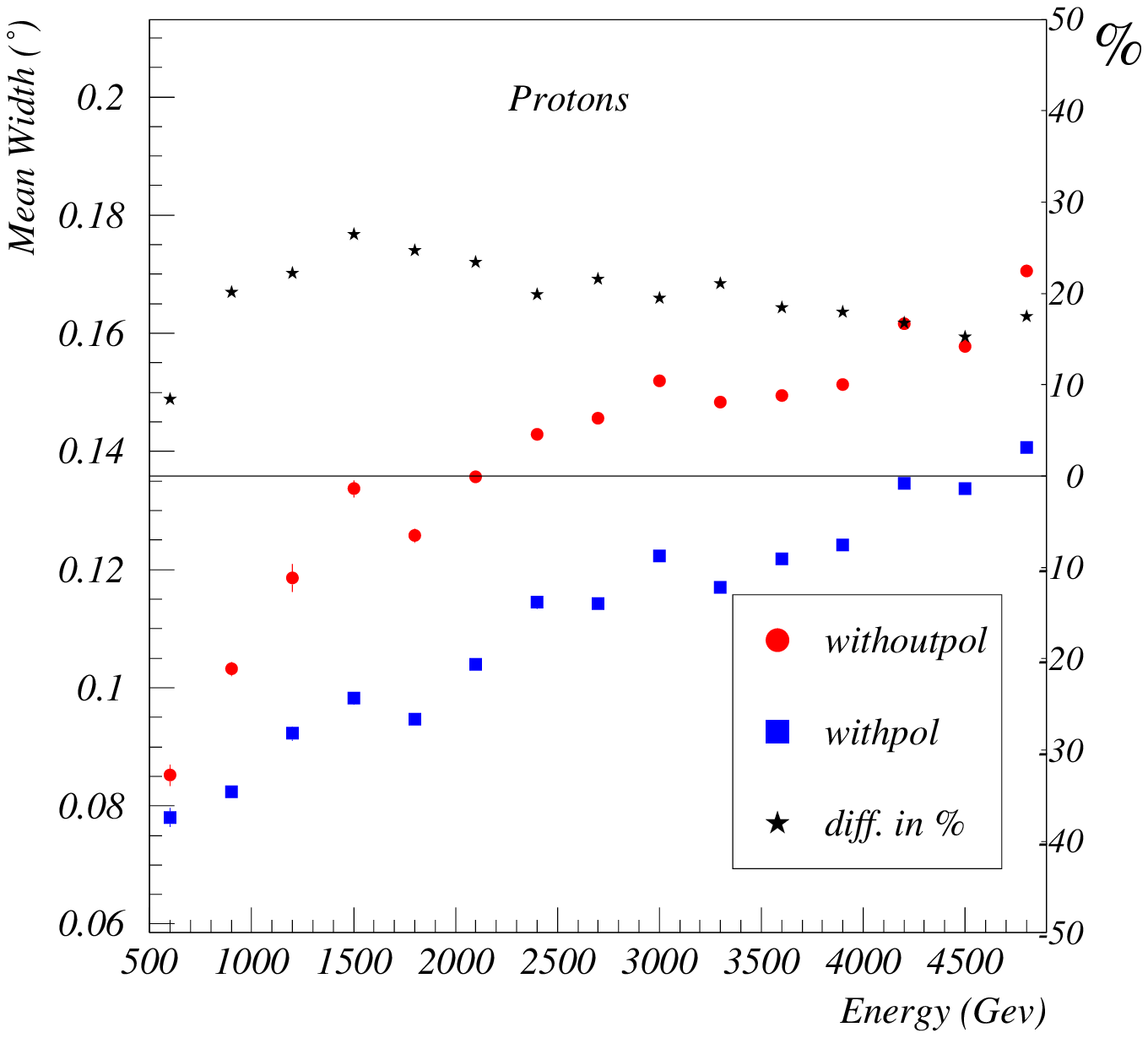,height=6.5cm,width=0.49\linewidth}\\
\caption{\it Comparison of the Hillas image parameters {\it length} (right) 
and {\it width} (left)
in the standard and the polarizers camera, for $\gamma$s and protons.
The absolute values are marked with circles and squares, while 
stars represent the percentage relative difference.}\label{imgpar}
\end{center}
\end{figure}

Plots in figure \ref{imgpar}, described in more detail in the caption,  
compare the mean value of the image parameters {\it length} and  
{\it width} for $\gamma$ and proton initiated EAS as a function 
of energy for both cameras. The general trend is a reduction in 
both parameters. It reflects the fact that the photons in the tails 
of angular distributions show a lower radial polarization.

It has been found that the distribution of the Hillas
$\alpha$ parameter is nearly unmodified. This is because
in spite of the fact that shower images in the polarizer 
arrangement are narrower, the shower images also become shorter
by approximately the same factor.

The effect of the polarizers on the quality factor for a standard set of
super-cuts~\cite[D. Petry et al. 1996]{mrk421} has been also studied. Results
indicate that the performance of the system is very similar
with and without polarizers, depending on the energy range.
However, given the approximations assumed in this work  
the result can not be considered a final answer. A detailed study
for each given instrumental setup would be necessary.

%----------------------------------
%    Figure  11
%----------------------------------

\subsection{High noise scenario}
\label{subsect53}

The previous section shows that the introduction of polarizers 
does not noticeably improve the performance of a telescope, although 
the signal to noise ratio increases. 
Notwithstanding this fact the arrangement could still be helpful 
when the amount of light arriving at the detector has to be reduced
to be able to cope with an increased background level.
Specific examples may be the need to adapt an instrument to be able 
to operate during twilight or in the presence of moon light.

\par

Observations during twilight, or in the presence of the moon enable
IACTs to extend their reduced duty cycle. The benefit lies not only
in the increased event statistics and observation time, but also in the
flexibility to efficiently cover the short term flux variations of 
VHE sources. Of course, there is a price to pay in the form of higher
background.

 Attempts to extend observation periods to high background conditions
as observations with moonlight have consisted either in using 
special UV sensitive PMTs  and blocking filters (since UV moonlight is 
blocked by the ozone layer) 
\cite[Pare et al 1991]{Pare},\cite[Chantell et al 1995]{Chantell},
\cite[Bradbury et al 1996]{Bradbury} 
\cite[D. Pomar\`ede et al. 2001]{moonwhipple}
or reducing
the PMT gain \cite[D. Kranich et al. 1999]{Kranich}.
Both approaches lead to an increase of the energy threshold of
the telescope, the last alternative having yielded the best results
to date.

Moonlight increases the amount of NSB by a factor of 3 to 5 during
half moon up to a factor 30-50 during full moon 
\cite[Dawson and Smith 1996]{Dawson} (when the telescope points 
45$^\circ$ away from the moon). 
In this section we will only consider situations where the background
light is unpolarized, leaving for section~\ref{sect6} the discussion
on polarized backgrounds.

To test the suitability of introducing polarizers during high background
observations three different setups have been contemplated:
\begin{enumerate}
\item[A] {\it Reduced PMT gain in the standard camera.}
\item[B] {\it Introduction of a grey filter on top of the standard camera,
         modeled by suppressing 50\% of the collected photons.}
\item[C] {\it Introduction of a polarizer camera.}
\end{enumerate}

 Setup {\bf B }, has been chosen to represent the easiest
solution of just blocking a fixed
percentage of the light arriving to the camera. Filters
tried in IACTs are smarter~\cite[D. Pomar\`ede et al. 2001]{moonwhipple}
, selecting a range of
wavelengths, but equally unrestricted in polarizations.

We have increased the NSB in our MC simulations in a factor 10.
This factor is a realistic estimate of a real high background 
situation, such as a moonlight observation.
The difference in noise among the three setups studied impose 
the need to tune first some parameters of the simulation. These are in 
particular $q_0$ (the trigger threshold for each pixel) and the two 
free parameters involved in the two level tail cut. 
The three of them have been adjusted to result on equal 
probabilities of accidental trigger, for the trigger,
and random occurrence, for the two level cut, respectively. 
In practice, as we are dealing with integer numbers, 
probabilities are only approximately equal. 
It has to be kept in mind that setups {\bf B } and {\bf C } 
reduce NSB to the same level,
leading to the same number of random triggers.

\begin{figure}[p]
\begin{center}
\epsfig{file=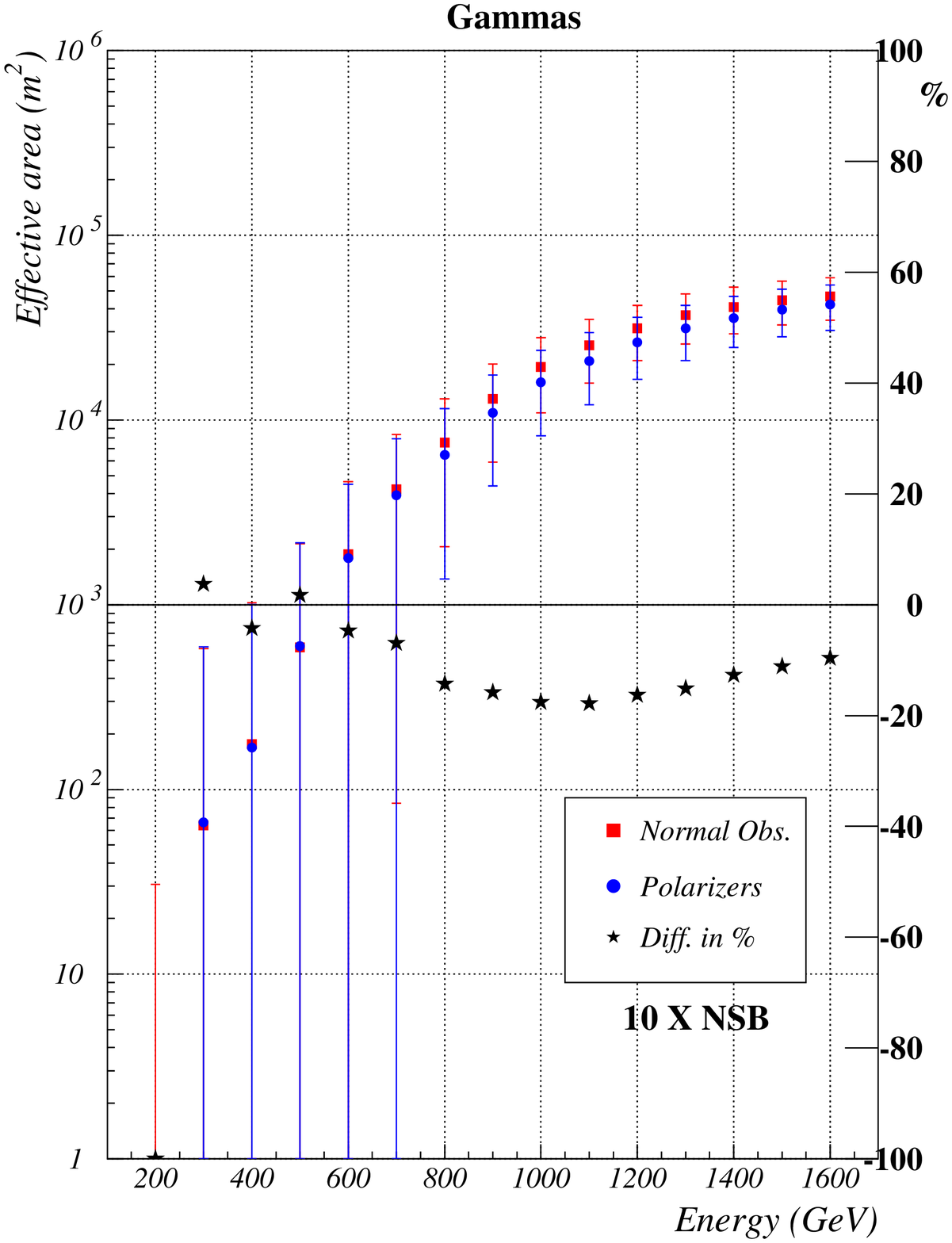,height=8cm,width=0.48\linewidth}%
\epsfig{file=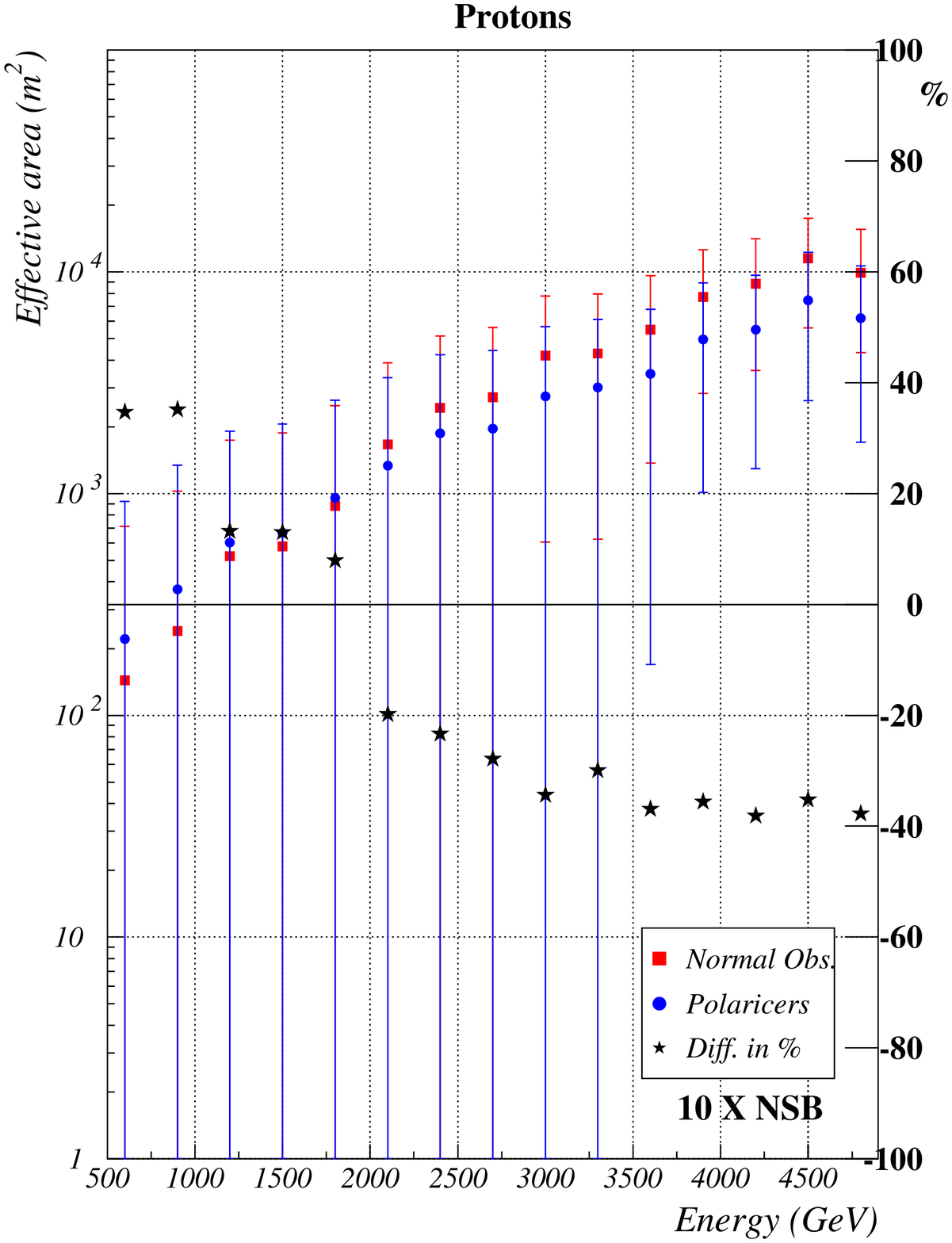,height=8cm,width=0.48\linewidth}\\
\caption{\it  Effective areas as a function of energy
for $\gamma$ and proton showers in
setups A and C under ten times the NSB of a dark night.
The right scale measures the percentage difference between 
the results obtained by both methods.}\label{10x-1}
\end{center}
\end{figure}

The results of our simulations can be summarized in two plots, 
showing the corresponding effective areas. 
Plot~\ref{10x-1} represents the effectives areas as a function
of energy when NSB is increased by a factor of 10 for case A
and case C. The reductions in the total 
amount of events observed are larger than those shown previously 
due to the different thresholds, but the conclusions are very similar to 
those in the previous section: polarizers suppress 
more light in proton showers but differences do not seem important 
enough to justify its use.

%----------------------------------
%    Figure  12
%----------------------------------

\begin{figure}[p]
\begin{center}
\epsfig{file=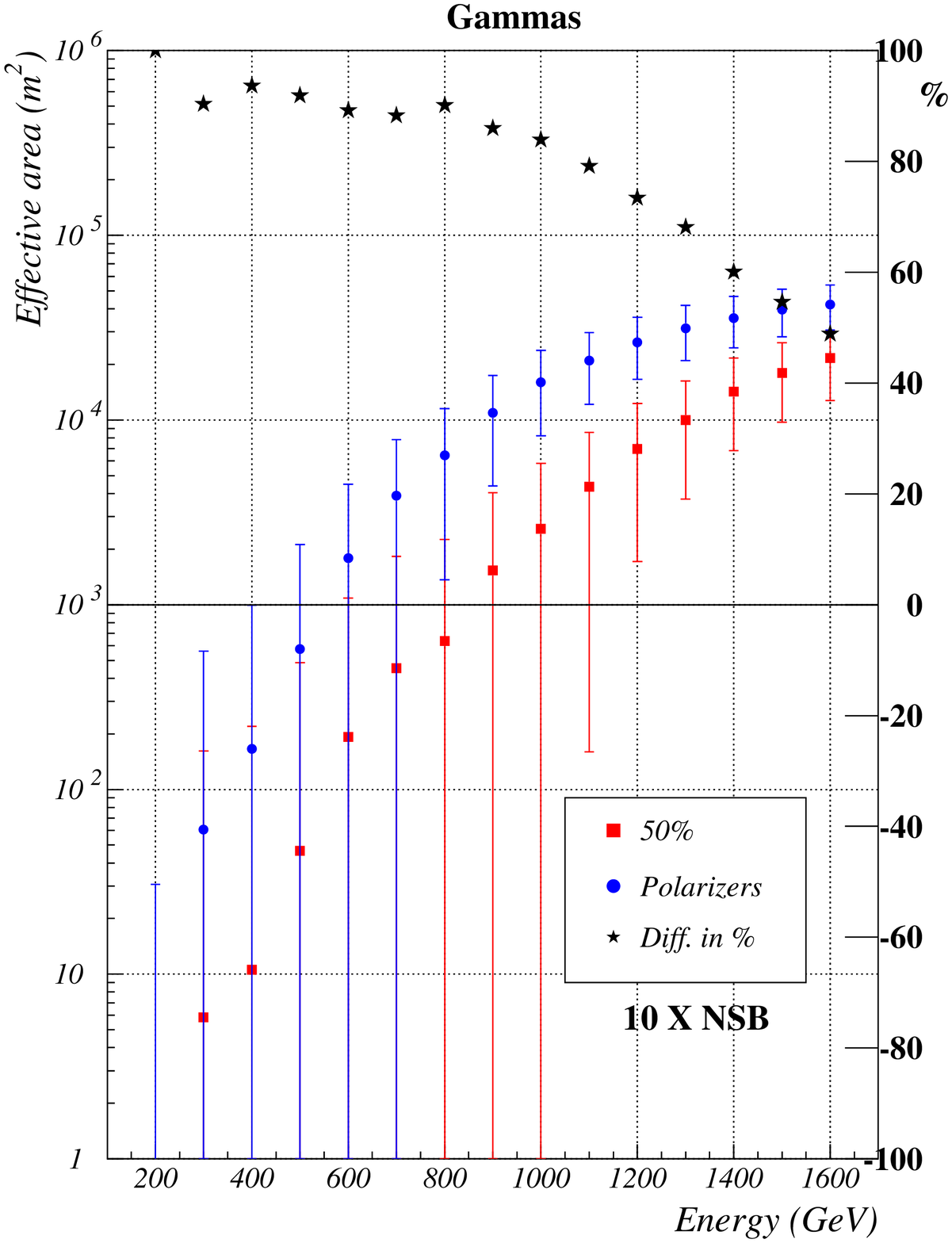,height=8cm,width=0.48\linewidth}%
\epsfig{file=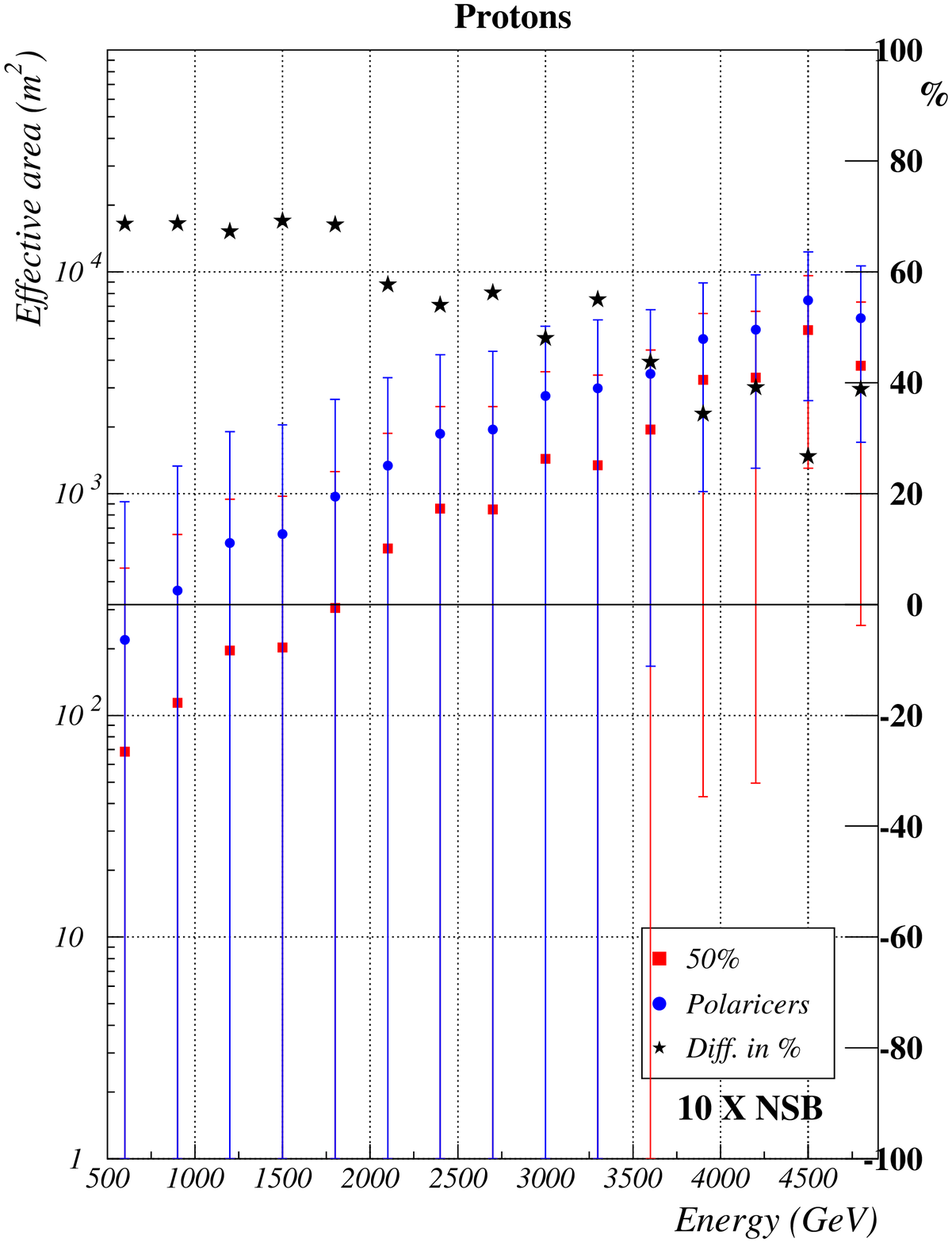,height=8cm,width=0.48\linewidth}\\
\caption{\it  Effective areas for $\gamma$ and proton showers for two
setups aiming at reducing the effect of NSB: light suppression using
a filter (labelled 50\%) and insertion of polarizers. 
The scale on the right axis measures the percentage difference 
between the results obtained in both setups.}\label{10x-2}
\end{center}
\end{figure}

Plot \ref{10x-2} compares the effect of polarizers and
of simple light suppression.
(labelled in the plots with {\it 50\%}).
From the left hand side plot it can be seen that the 
number of surviving gammas
is higher for the polarizers case by a factor that 
goes from 50 to 100\%. The right hand
side plot shows that more protons survive,
and the gain for protons is always
smaller. Even an equal increase in protons and gammas 
would favor the use of polarizers, 
since the signal grows linearly with the number
of events in the asymptotic Gaussian regime, while 
background grows only as the square root. 
The fact that the fraction is
larger for gammas makes it more attractive.

%----------------------------------
%    Figure  12
%----------------------------------

\section{Observations with polarized background light}
\label{sect6}

While the considerations made in previous sections apply 
to unpolarized background light,
attention must be paid to situations where the background 
has a definite polarization.
The question of whether the use of polarizers could help 
in those situations
can be partially answered from the results shown so far.

The most common case where polarized light constitutes an appreciable 
background is the one of linear polarization, normally  caused by scattering.
An important case is that of observation in  
moonlit nights, at a certain angular
separation from the moon or at twilight, away from the sun.
Light scattered at right angles to the moon or sun position 
acquires a high degree of polarization. 
See for example \cite[I.K. Baldry and E. Bland-Hawthorn]{moonpol}
and references therein. 
At least two simple arrangements of polarizers can be imagined as possible 
improvements for this situation. The first one is the polarizer camera 
discussed in the paper.
The second one consists in a single polarizer whose transmitting axis
is perpendicular to the dominant polarization of the background, 
to block a maximum amount of it.

Besides the reduction in background
(which may be very important under some circumstances),
the most important effect of both arrangements
is the creation of "non-uniformity" on the camera.
The fraction
of the polarizers aligned parallel to the direction of the background
allows light to pass freely, increasing the background with respect to
those perpendicular to it. 
A single polarizer oriented perpendicularly
to the background polarization will suppress 
the light more efficiently depending on the position of the image 
on the camera (as discussed in section ~\ref{sect2}). Both effects 
should be taken into account in designing such a system.

\section{Conclusions}
\label{sect7}

We hope that this paper will help to clarify some of the
properties of Cherenkov light polarization in EAS. 
We have pinpointed several aspects of the Cherenkov
detection which may benefit from the consideration
of the state of the light polarization. 
In particular we have considered $\gamma$/hadron separation, 
reduction of NSB and reduction of the trigger gate width,
with its implications a further disminution of the NSB.

More specific results have been presented for the special case of IACTs.
An experimental setup which uses the polarization to increase
the significance of VHE gamma signals has been suggested.
In this setup polarizing filters are arranged on top of 
each of the camera PMTs. The
transmitting axes point to the camera center.
We have studied the performance of this setup by means of a simplified MC
simulation.  
Our conclusion is that the setup does not help in dark night 
observations but would be useful when the high NSB makes 
necessary to reduce the amount of light on the camera. 
In that case it amounts to an {\it intelligent} reduction of light. 

Special attention must be paid to situations when the background
light is polarized.

In this study some factors (such as the characteristics of 
the telescope, the composition of the hadronic background or the 
behavior of the polarizers) have been greatly simplified. 
At the same time the analysis has not been
refined to fully exploit all of the advantages of the polarizer camera, 
as for instance by taking into account the shorter time profile of the 
Cherenkov photons or by redefining the set of super-cuts. 
A more detailed study featuring all the characteristics of 
a real detector and including experimental tests may prove rewarding.

\section*{Acknowledgements}
\label{sect8}

The authors are indebted to many members of the HEGRA collaboration for
fruitful discussions held during the development of this study.  I. de la
Calle wishes to thank the High Energy Astrophysics group of the University of
Leeds for their suggestions and comments.

This work has been supported by the Spanish funding agency CICYT under
contract AEN98-1094.

%\documentclass{article}

%\usepackage{epsfig}
%\usepackage{multirow}

%\usepackage[nofiglist,notablist,nomarkers]{endfloat}

%\def\deg{\ensuremath{^{\circ}}}
%\renewcommand{\u}[1]{\ensuremath{\mathrm{#1}}}

%\begin{document}

%
%
%
%\newpage
%
%
%
%\newpage
%
%
%
%\newpage
%
%
%
%
%
%
%\newpage
%
%
%
%
%
%
%\newpage
%
%
%
%
%
%
%\newpage
%
%
%
%
%
%
%\newpage
%
%
%
%
%
%
%\newpage
%
%
%
%
%
%
%\newpage
%
%
%
%
%
%
%\newpage
%
%
%
%
%
%
%\newpage
%
%
%
%
%
%
%\newpage
%
%

%\end{document}

%\documentclass{article}

%\usepackage{epsfig}
%\usepackage{multirow}

%\usepackage[nofiglist,notablist,nomarkers]{endfloat}

%\def\deg{\ensuremath{^{\circ}}}
%\renewcommand{\u}[1]{\ensuremath{\mathrm{#1}}}

%\begin{document}

\begin{table}[h]
\large
\begin{center}
\begin{tabular}{|c||c|c|c|c|c|}   
\hline
 & 
\multicolumn{2}{c|} {Low NSB (dark night)} & 
\multicolumn{3}{c|} {High NSB} \\ 
  
{\small{$FLUX_{NSB}$}} &
\multicolumn{2}{c|} {\small{$(1.7 \pm 0.4) \cdot 10^{12}$ }} & 
\multicolumn{3}{c|} {\small{$(1.7 \pm 0.4) \cdot 10^{13}$ }} \\ 

{\small{ph m$^{-2}$ sr$^{-1}$ s$^{-1}$}} & 
Without & 
With    & 
Without & 
With    & 
50\% \\

\hline 
$\ge$ 2NN/271 & 
10 phe      & 
 8 phe      & 
26 phe      & 
19 phe      & 
19 phe      \\ 
\hline
\end{tabular} 
\caption{Trigger threshold conditions for the two
NSB conditions and three experimental setups described in the text
(camera with and without polarizers and 50\% light suppression) 
\label{Tab1}}
\end{center} 
\end{table}

%\end{document}

\end{document}